\documentclass[usenatbib]{mnras}

\usepackage{multirow}
\usepackage{rotating}
\usepackage{colortbl}
\usepackage{color}
\usepackage[fleqn]{amsmath}
\usepackage{amssymb}
\usepackage{amsfonts}
\usepackage{verbatim}
\usepackage{scalefnt}
\usepackage[percent]{overpic}
\usepackage[T1]{fontenc} 
\usepackage{aecompl} 
\usepackage{ulem}
\usepackage{cleveref}

\def\lsim{\mathrel{\rlap{\lower 3pt \hbox{$\sim$}} \raise 2.0pt \hbox{$<$}}}
\def\gsim{\mathrel{\rlap{\lower 3pt \hbox{$\sim$}} \raise 2.0pt \hbox{$>$}}}

\newcommand{\comments}[1]{} 
\newcommand\T{\rule{0pt}{2.6ex}}       
\newcommand\B{\rule[-1.2ex]{0pt}{0pt}} 

\newcommand{\soutPC}{\bgroup\markoverwith{\textcolor{cyan}{\rule[0.5ex]{2pt}{1pt}}}\ULon}

\title[Introducing the \textsc{Phoebos} simulation]{Introducing the \textsc{Phoebos} simulation: galaxy properties at the dawn of galaxy formation}

\author[F. van Donkelaar et al.] {Floor van Donkelaar,$^{1}$\thanks{E-mail: floor.vandonkelaar@uzh.ch} Pedro R. Capelo,$^{1}$ Lucio Mayer,$^{1}$ Darren S. Reed$^{1,2}$ and Thomas R. Quinn$^{3}$
\\
$^{1}$Department of Astrophysics, University of Zurich, Winterthurerstrasse 190, CH-8057 Z{\"u}rich, Switzerland\\
$^{2}$Science IT, University of Zurich, Winterthurerstrasse 190, CH-8057 Z{\"u}rich, Switzerland\\
$^{3}$Astronomy Department, University of Washington, Seattle, WA 98195, USA}

\date{Accepted XXX. Received YYY; in original form ZZZ}

\pubyear{2025}

\begin{document}

\label{firstpage}

\pagerange{\pageref{firstpage}--\pageref{lastpage}}

\maketitle


\begin{abstract}
The James Webb Space Telescope (JWST) now allows us to observe galaxies at the end of cosmic dawn ($z \sim 10$--15) with unprecedented detail, revealing their morphologies, sizes, and internal structures. These observations offer crucial insights into the physical processes driving early galaxy formation. In this work, we introduce the \textsc{Phoebos} hydrodynamical cosmological simulation, a state-of-the-art 100~Mpc volume designed to study the formation and evolution of galaxies at the end of cosmic dawn and into the epoch of reionization. \textsc{Phoebos} includes a stellar feedback model that is intentionally weak, in order to address the high abundance of massive galaxies seen by JWST at early epochs. At variance with most large cosmological hydrodynamical simulations, we do not employ an effective equation of state model, instead our radiative cooling model allows us to capture the multi-phase nature of the gas inside and around galaxies. \textsc{Phoebos} reproduces key observables of early galaxy formation at $z \gtrsim 8$, including the stellar mass function and the stellar-to-halo mass relation. It also recovers the observed slope of the stellar size-to-mass relation and matches the specific star formation rate remarkably well. These results suggest that highly efficient star formation in the presence of only mild regulation from stellar feedback, drives early galaxy growth, supporting a scenario of rapid stellar mass assembly during cosmic dawn. There are indications in the cosmic star formation density that, at lower redshifts, \textsc{Phoebos} might overpredict the stellar mass within the systems, suggesting that a transition to a stronger stellar feedback may be necessary to reproduce later-time observations. These results highlight the potential of \textsc{Phoebos} to interpret JWST observations and to probe the evolving physical processes that shape galaxy formation.
\end{abstract}



\section{Introduction}\label{sec:introduction}
Since the earliest observations of galaxies, their extended structures have provided key clues about their unique properties and complex formation histories. From the Hubble Space Telescope \citep[HST; e.g. ][]{Petersen:1995} to the latest advances with the James Webb Space Telescope \citep[JWST;][]{JWST}, studying the diversity of galaxies has remained a major focus in astronomy. Now, with JWST, we can explore the earliest galaxies in unprecedented detail, revealing their structures, shapes, and sizes up to $ z \sim 10$ \citep[e.g.][]{Ferreira:2022aa, Harikane:2023aa, Huertas:2023aa, Kartaltepe:2023aa, Ono:2023aa, Adams:2024aa, Ito:2024aa, Ormerod:2024, yang:2025aa, Shuntov:2025aa}. For the first time, we can directly examine the stellar structures of galaxies at cosmic dawn, providing new insights into the physics that shaped their formation.

Because the same physical processes that drive galaxy formation, such as mergers, accretion, and feedback, also determine their sizes, studying how sizes vary across a population reveals their growth histories over cosmic time. One of the most fundamental ways to probe these histories is through galaxy morphology, which reflects both their internal dynamics and their interactions with the surrounding environment, including the influence of their host dark matter (DM) haloes \citep[e.g.][]{Bezanson:2009aa, vanderwel:2014aa}. Previous HST studies have shown that the slope of the stellar size-to-mass relation remains largely unchanged from $z \sim 0$ to $z \sim 3$ \citep[e.g.][]{vanderwel:2014aa}. Additionally, galaxy size distributions have been found to follow a log-normal form with stable scatter up to $z \sim 8$ \citep[e.g.][]{Shibuya:2015aa, Kawamata:2015aa, Kawamata:2018aa}. However, key questions remain open, such as when the first galactic discs emerged and how feedback mechanisms influenced early galaxy sizes. Observations with the Atacama Large Millimeter/submillimeter Array \citep[ALMA;][]{ALMA} have revealed a significant fraction of dynamically cold star-forming discs as early as $z \sim 6.8$ \citep[e.g.][]{Smit:2018aa, Rizzo:2020aa, Rizzo:2021aa, Roman:2023aa}, a finding that aligns with predictions from simulations \citep[e.g.][]{Tamfal:2022aa, Michael:2022aa, Kohandel:2023aa, Donkelaar:2025aa}.

There is a broad consensus, from both space- and ground-based observations, that galaxies have undergone strong size evolution since $z \sim 3$, with galaxies at $z \sim 2$ being a factor of 2--7 smaller than their local counterparts of similar mass  \citep[e.g.][]{Daddi:2005aa, Trujillo:2007aa, Buitrago:2008aa, Bezanson:2009aa, vandokkum:2010aa, Barrow:2017aa, Koekemoer:2011aa, vanderwel:2014aa, Mowla:2019aa, Suess:2019aa, Mosleh:2020}. For example, quiescent galaxies with stellar masses around $\log(M_*/{\rm M}_{\sun}) \sim 11$ had effective radii of $R_{\rm eff} \sim 1.0$--$1.8$~kpc\footnote{In this work, comoving length units have the `c' prefix, whereas physical length units do not have any prefix.} at $z \sim 2$ \citep{vandokkum:2008aa, Buitrago:2008aa, vanderwel:2014aa}, compared to $R_{\rm eff} \sim 5$--$7$~kpc locally. Star-forming galaxies at $ z \sim 2$ were somewhat larger than their quiescent counterparts, with typical sizes of $R_{\rm eff} \sim 3.5$--$4.5$~kpc \citep{vanderwel:2014aa, Allen:2017aa}, yet still more compact than similar galaxies in the local Universe \citep{Mowla:2019aa, Suess:2019aa}. 

In particular, JWST has revealed a population of extremely compact galaxies at $z \gtrsim 6$ \citep[e.g.][]{Baggen:2023aa, Morishita:2024aa}. For example, \citet{Miller:2024aa} have shown that the sizes of galaxies with $\log (M_*/$M$_{\sun}) = 8.5$ grow rapidly, from an average of 400~pc at $z = 8$ to 830~pc at $z = 4$. This evolution is much faster than what expected from power-law scalings using HST observational data, which could accurately described galaxy size growth at $z < 2$ in previous studies. \citep[e.g.][]{Lange:2015aa, carlsten:2021aa, nedkova:2021aa}. This discrepancy could indicate that such relations break down at higher redshifts and suggest that low-mass galaxies during the epoch of reionization are thus shaped by different physical processes, potentially stronger in the sense of more efficient star formation than those operating at later cosmic times. 

Moreover, recent analyses using JWST’s Near Infrared Camera imaging from the Cosmic Evolution Early Release Science Survey have significantly advanced our understanding of galaxy sizes at high redshift, particularly for $2 < z < 10$. For example, \citet{yang:2025aa} measured nearly 1900 galaxies, showing that sizes decline with increasing redshift, roughly following a $(1 + z)^{-1.1}$ scaling. This is consistent with theoretical expectations from DM halo evolution \citep[e.g.][]{Mo:1998aa} and earlier HST studies \citep[e.g.][]{vanderwel:2014aa, Shibuya:2015aa}. Nevertheless, JWST’s optical resolution reveals significant scatter in galaxy sizes at fixed stellar mass or ultraviolet (UV) luminosity, with a diverse range of morphologies, including ultra-compact and extended systems, already at $z \gtrsim 6$ \citep[e.g.][]{Baggen:2023aa, Morishita:2024aa, Miller:2024aa, yang:2025aa}. 

Concurrently, studies have found that while the slope and scatter of the size-mass/luminosity relation remain stable across $0 \leq z \lesssim 8$, the surface density of the star formation rate (SFR) increases significantly toward higher redshifts \citep[e.g.][see also, e.g. \citealt{Kannan:2025aa} for results from simulations]{Oesch:2010aa, Bruce:2012aa, Mosleh:2012aa, Ono:2013aa, vanderwel:2014aa, Morishita:2014aa, Shibuya:2015aa, Allen:2017aa, Mosleh:2020, Bouwens:2022aa}. Early JWST results have furthermore reinforced the idea that discs formed surprisingly early, with galaxy sizes at $z \geq 10$ following well-defined trends, and some showing compact disc-like structures \citep[e.g.][]{Ferreira:2022aa, Robertson:2023aa, Ward:2024aa, Xu:2024aa}. Despite these advancements, most of the findings have focused on galaxies at $z < 10$, making statistical investigations of even earlier galaxies an important next step to take. Comparing results with theoretical models will be crucial for our understanding of the fundamental processes driving galaxy formation in the first billion years after the Big Bang \citep[e.g.][]{Donkelaar:2022aa, Costantin:2024aa, Yajima:2023aa, Shen:2024aa, nakazato:2024aa, Punyasheel:2024aa, LaChance:2024aa, Vega:2024aa, Donkelaar:2025aa, Mayer:2025aa, Burger:2025aa, Kannan:2025aa}

Furthermore, JWST’s detection of an unexpectedly large number of ultra-high-redshift ($z \gtrsim 10$--15) galaxies with unexpectedly high stellar masses has sparked intense debate regarding its implications for the $\Lambda$CDM model, which predicts significantly fewer massive haloes at such early epochs \citep[e.g.][]{Robertson:2024aa}. While there have been claims that these observations are in fundamental tension with standard $\Lambda$CDM \citep[e.g.][]{Boylan:2023aa, lovell:2023aa}, the situation is complicated by various uncertainties. These include the precise mass of DM haloes, the efficiency of early SF, and the contribution of UV light from star-forming regions. Potential explanations put forward to reconcile the observations include enhanced SF at early times \citep[][]{Dekel:2023aa, Li:2024aa, somerville:2025aa, yang:2025aa}, reduced dust attenuation \citep{Ferrara:2023aa}, the presence of top-heavy initial mass functions \citep{Trinca:2024aa, Yung:2024aa}, and even contributions from accreting supermassive black holes \citep{Trinca:2024aa}.

\begin{table*}
\centering
\begin{tabular}{c|ccccccc}
\hline
Run:       & \# DM    & \# gas   & \# tot & $m_{\rm DM}$ [M$_{\sun}$] & $m_{\rm gas}$ [M$_{\sun}$] & $\epsilon$ [kpc] \T \B \\ \hline
\textsc{Phoebos}   & $2904^3$ & $1944^3$ & $3.184 \times 10^{10}$ & $1.360 \times 10^6$ & $8.473 \times 10^5$ & 0.30 \T \B \\
\textsc{PhoebosLR}   & $1452^3$ & $972^3$  & $3.980 \times 10^{9}$  & $1.088 \times 10^7$ & $6.778 \times 10^6$ & 0.60 \T \B \\
\textsc{PhoebosULR}  & $726^3$  & $486^3$  & $4.974 \times 10^{8}$  & $8.701 \times 10^7$ & $5.423 \times 10^7$ & 1.20 \T \B \\
\end{tabular}
\caption{Specifications for the three runs described in this work. The quoted softening is valid for $z < 9$. For $z \ge 9$, we use $10 \epsilon/ (1+z)$. All simulations were run in a [100~cMpc]$^3$ volume, assuming the Planck 2018 cosmology.
}
\label{tab:resolution}
\end{table*}

To fully understand the physical processes driving early galaxy evolution, it is crucial to compare the most recent high-redshift observations with predictions from state-of-the-art galaxy formation models. Numerical simulations provide a fundamental framework for making sense of observed galaxy properties, offering a way to test theoretical predictions directly against real data. These galaxy formation models do not only refine our understanding of structure formation but also improve the accuracy of observational techniques and guide future observational survey strategies \citep[e.g.][]{Pforr:2012aa, Pforr:2013aa, Smith:2015aa}. This is particularly important at high redshift, where observations are still limited. 

Over the past decade, numerical simulations have provided critical theoretical insights into the sizes and morphologies of galaxies in the early Universe. The Feedback in Realistic Environments (FIRE) project \citep[][]{Hopkins:2014aa,Hopkins:2018aa}, for instance, has developed cosmological simulations that directly resolve the interstellar medium (ISM) of individual galaxies while capturing their cosmological environments. Notably, \citet{Ma:2018aa} utilized the FIRE-2 simulations \citep{Ma:2018ab} to study galaxy morphologies and sizes at $z>5$, focusing on the low-mass end. Similarly, \citet{Marshall:2022aa} examined galaxy sizes at $7 < z < 12$ using mock images generated from the Bluetides simulations \citep{Wilkins:2016aa, Wilkins:2017aa}, emphasizing the significant role of dust. \citet{Roper:2022aa} explored galaxy sizes at $z \geq 5$ using the FLARES simulations \citep{Lovell:2021aa} and investigated the physical mechanisms driving compact galaxy formation \citep{Roper:2023aa}.

Moreover, \citet{Shen:2024aa} identified a correlation between the intrinsic half-mass radius and stellar mass in the lower mass range of galaxies, utilizing the \textsc{Thesan} simulation \citep[][]{Kannan:2022aa}. Their work revealed a dual-phase trend, showing that the sizes of low-mass galaxies are strongly influenced by stellar feedback-driven outflows. While the FLAMINGO project \citep[][]{Schaye_et_al_2023} focuses on a different aspect, providing cosmological hydrodynamical simulations for large-scale structure and galaxy cluster surveys, it similarly contributes to our understanding of galaxy formation and evolution. The simulations reproduce key observables such as the galaxy stellar mass function and cluster gas fractions  at low redshift ($z \lesssim 1$), and show that baryonic effects suppress structure on small scales. Other notable numerical simulations that have examined the high-redshift Universe include, but are not limited to, \textsc{Vulcan} \citep{Anderson:2017aa}, which look into  the number densities of undetectable, faint galaxies and their escape fractions of ionizing radiation during reionization; the First Billion Years simulation suites \citep{Paardekooper:2013aa}, which explore early galaxy formation and reionization; the Renaissance simulation suite \citep{Xu:2016aa, Barrow:2017aa}, focusing on the formation of the first galaxies; the IllustrisTNG project \citep[e.g.][]{Nelson_et_al_2018, Nelson:2019aa}, investigating galaxy evolution and feedback processes across cosmic time; and the Sphinx simulation \citep{Rosdahl:2018aa}, which models radiation hydrodynamics to study the impact of radiation feedback on galaxy formation. 

In this paper, we introduce the \textsc{Phoebos} hydrodynamical cosmological volume simulation, a state-of-the-art numerical framework designed to explore the formation and evolution of galaxies in the early Universe.  The primary aim of this paper is twofold: first, to demonstrate the capability of \textsc{Phoebos} in accurately reproducing key galaxy statistics at high redshift; and second, to provide a comprehensive analysis of galaxy sizes at redshifts $z \gtrsim 8$ across a wide range of galaxy stellar masses.

What sets the \textsc{Phoebos} simulation apart from most other cosmological models is its unique interpretation of feedback, which emphasizes weaker feedback compared to many recent models, with no active galactic nucleus (AGN) feedback being included. Despite relying on more traditional stellar physics recipes, this alternative approach to feedback at high redshift might be able to help explain the unexpectedly large number of ultra-high-redshift galaxies detected by JWST. Moreover, it could provide a valuable baseline for assessing the role and efficacy of AGN feedback in galaxy formation models.

\textsc{Phoebos} solves the full energy equation, including all relevant atomic and molecular cooling lines using \textsc{Cloudy} tables \citep{Shen_et_al_2010,Shen_et_al_2013} in the presence of a time-dependent cosmic ionizing UV background. Cooling by hydrogen and helium is computed by solving directly the time-dependent rate equations, hence without assuming equilibrium, at variance with most cosmological volume simulations. It is important to note that most cosmological hydrodynamical volume simulations of comparable or larger scale, such as \textsc{TNG100} \citep[e.g.][]{Nelson_et_al_2018}, \textsc{Eagle} \citep[e.g.][]{Schaye:2015aa}, and \textsc{Thesan} \citep[e.g.][]{Kannan:2022aa}, do not explicitly solve for cooling and heating processes in the high-density gas phase within galaxies. Instead, they employ an effective equation of state \citep{Springel:2003aa}, which suppresses the multi-phase structure of the ISM to improve convergence with resolution. A markedly different impact on feedback of the two approaches is expected. Resolving the clumpy structure of the cold ISM, where stars form, is essential for refining galactic models and feedback mechanisms, ensuring a more realistic representation of SF and energy exchange in the very compact, dense gas discs expected in the early stages of galaxy formation.

The paper is structured as follows.  Section~\ref{sec:numerical_setup} introduces the simulation suite, outlining the numerical setup and physical models employed. In Section \ref{sec:results}, we present the main results of our simulations, focusing on the statistical properties of galaxies to assess the reliability of our framework. Additionally, we examine the size-to-mass relation of galaxies across different redshifts. Finally, Section \ref{sec:conclusions} summarizes our conclusions and discusses the implications of our findings.

\section{Numerical setup}\label{sec:numerical_setup}

The choice of our simulation parameters was informed by another cosmological simulation, \textsc{Romulus25} \citep[][]{Tremmel_et_al_2017}, which was run with the same code \citep[and very similar subgrid models, except for their inclusion of black hole physics and exclusion of high-temperature metal-line cooling; see][]{Tremmel_et_al_2019} used in this work (\textsc{ChaNGa}; \citealt{Jetley:2008aa,Jetley:2010aa,Menon:2015aa}) and has been successful at reproducing, e.g. the black hole and galaxy mass functions, the AGN luminosity function, and the black hole-galaxy scaling relations over a wide range of black hole and galaxy masses and redshifts.\footnote{We stress that the nature and impact of black hole physics at high redshift -- the target of our simulations -- is still unclear, hence we conservatively exclude it. Moreover, it was shown in, e.g. the \textsc{FIREBox} simulation \citep[][]{Feldmann_et_al_2023} that the mean galaxy population properties can be recovered also without the modelling of black holes.} In \textsc{Romulus25}, a total of $1152^3$ DM particles and $768^3$ gas particles were simulated, assuming a \citet{Planck_2013_cosmo_params} cosmology, in a volume of [25~cMpc]$^3$, with a resulting particle mass of $3.39 \times 10^5$~M$_{\sun}$ for DM and $2.12 \times 10^5$~M$_{\sun}$ for gas.

The relatively small volume of \textsc{Romulus25} does not allow investigating also the highest-density peaks, which is part of the scientific goals of this work. Therefore, we increased the volume 64-fold to $V=[100\, {\rm cMpc}]^3$, in order to have enough statistics also on the most massive structures in the Universe. Keeping the same mass resolution as that of \textsc{Romulus25} in such a large volume would be computationally very demanding. Thus, we kept a similar DM-to-gas particle number ratio (to decrease numerical noise) but considered instead $N_{\rm DM} = 2904^3$ DM particles and $N_{\rm gas} = 1944^3$ gas particles (i.e. a factor of sixteen more particles than in \textsc{Romulus25}). Assuming a more recent set of cosmological parameters \citep[Planck 2018; $\Omega_{\rm m} = 0.3111$, $\Omega_{\Lambda} = 0.6889$, $\Omega_{\rm b} = 0.0490$, $h = 0.6766$, $n_{\rm s} = 0.9665$, $\sigma_8 = 0.8102$;][]{Planck_2018_cosmo_params}, this yields particle masses of $m_{\rm DM} = 1.360 \times 10^6$~M$_{\sun}$ and $m_{\rm gas} = 8.473 \times 10^5$~M$_{\sun}$ (i.e. just a factor of four more massive than in \textsc{Romulus25}, in a much larger volume).

For both DM and baryons, we assumed a spline force softening $\epsilon = 0.300$~kpc (i.e. a Plummer equivalent force softening of 0.214~kpc; see \citealt{Kim_et_al_2016}) for $z < 9$ and $\epsilon = 3.0/(1+z)$~kpc for $z \ge 9$, and a minimum smoothing length equal to 5 per cent of the softening.\footnote{We note that these values are slightly different from those of \textsc{Romulus25}, wherein the spline force softening was 0.350~kpc for $z < 8$ and $3.15/(1+z)$~kpc for $z \ge 8$, and the minimum smoothing length was 20 per cent of the softening. In the redshift range of interest ($8 \le z \lesssim 15$), the minimum softening is $\sim$0.2~kpc for both simulations.}

\begin{figure}
    \centering
    \includegraphics[ trim={0cm 0cm 0cm 0cm}, clip, width=0.48\textwidth, keepaspectratio]{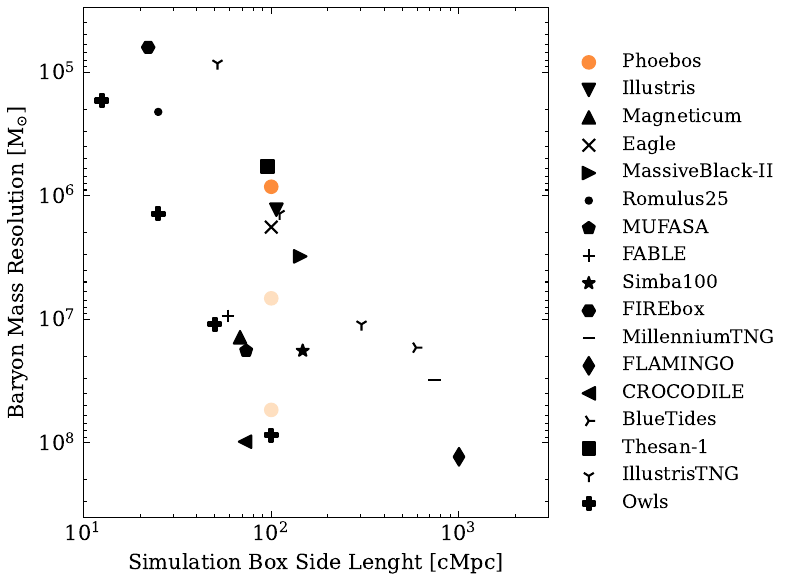}
    \caption{Baryonic mass resolution and box size for several cosmological hydrodynamical simulations from the recent literature: \textsc{Illustris} \citep[e.g.][]{Sijacki:2015aa}, \textsc{Magneticum} \citep[][]{Hirschmann:2014aa}, \textsc{Eagle} \citep[][]{Schaye:2015aa}, \textsc{Romulus25} \citep[][]{Tremmel_et_al_2017}, \textsc{MUFASA} \citep[][]{Dave:2016aa}, \textsc{FABLE} \citep[][]{Henden:2018aa}, \textsc{Simba100} \citep[][]{Dave:2019aa}, \textsc{FIREbox} \citep[][]{Feldmann_et_al_2023}, MilleniumTNG \citep[e.g.][]{Hernandez:2023aa}, FLAMINGO \citep[][]{Schaye_et_al_2023}, \textsc{CROCODILE} \citep[][]{Oku:2024aa}, \textsc{BlueTides} \citep[][]{Feng:2015aa}, \textsc{Thesan}-1 \citep[][]{Kannan:2025aa}, IllustrisTNG \citep[e.g.][]{Nelson:2019aa, Nelson_et_al_2018} and \textsc{OWLS} \citep[][]{Schaye:2010aa}. Filled orange circles correspond to the three simulations presented in this work. We compare these to a selection of cosmological hydrodynamical simulations from the literature that include radiative cooling, have a volume of at least [10~cMpc]$^3$, and run down to $z = 0$, with exceptions for simulations focusing on the early Universe: \textsc{BlueTides}, which ends at $z \sim 7$; \textsc{Thesan}-1, which ends at $z = 5.5$, and the highest-resolution \textsc{Phoebos} run, which ends at $z \sim 8$.}
    \label{fig:comparison}
\end{figure}

We also ran two lower-resolution simulations -- \textsc{PhoebosLR} and \textsc{PhoebosULR} -- in which, at each lower-resolution level, we multiplied the particle masses by eight and the particle softenings by two, and divided the particle number by eight (see Table~\ref{tab:resolution}). In Figure~\ref{fig:comparison}, we show the baryonic mass resolution and the cosmological volume of our three simulations, compared with other cosmological simulations from the literature.

The initial conditions (ICs) were built using \textsc{music2} \citep[][]{Hahn_et_al_2021,Rampf_et_al_2021}, a cosmological IC generator that improves upon older generators \citep[e.g. \textsc{music};][]{Hahn_Abel_2011} by providing higher-order Lagrangian perturbation theory (LPT) ICs for two gravitationally coupled fluids (e.g. DM and gas) in the cold limit (i.e. neglecting finite temperature perturbative effects).\footnote{Since \textsc{music2} does not currently handle different numbers of particles, we had to first create two IC files (e.g. one file with $2904^3$ DM particles and $2904^3$ gas particles and one file with $1944^3$ DM particles and $1944^3$ gas particles, for the high-resolution simulation), remove the DM particles from the $1944^3$ file and the gas particles from the $2904^3$ file, and merge the two resulting files.} We generated a Gaussian random realization of initial density fluctuations whose power spectrum is described by the  \citet{Eisenstein_Hu_1998} transfer function. Beginning with a regular three-dimensional grid, the density field was converted into particle displacements. For our particular case, we used the third-order LPT (3LPT) and a simple cubic particle load (equal numbers of particles and grid cells), with an initial redshift\footnote{The initial redshift was chosen because we wish to start analyzing at $z \sim 25$ and we used the ``factor of ten'' rule of thumb (assuming 2LPT) for the ratio between initial and ``analysis'' redshift \citep[][]{Reed:2013aa}. This should be conservative, given our choice of 3LPT ICs, which suffer less from errors due to truncation of higher-order terms \citep[][]{Michaux:2021aa} -- i.e. `transients', which are largest at early simulation times \citep[][]{Scoccimarro:1998aa, Crocce:2006aa}.} of $z_{\rm init} = 249$.  Gas was initialized with temperature and metallicity equal to $T_{\rm init} = T_{\rm CMB,0}(1+z_{\rm init}) = 681.5$~K and zero, respectively, where $T_{\rm CMB,0}$ is the average cosmic microwave background (CMB) temperature at $z = 0$, and assuming no baryon streaming motions relative to DM.

\begin{figure*}
    \centering
    \includegraphics[ trim={0cm 0cm 0cm 0cm}, clip, width=1\textwidth, keepaspectratio]{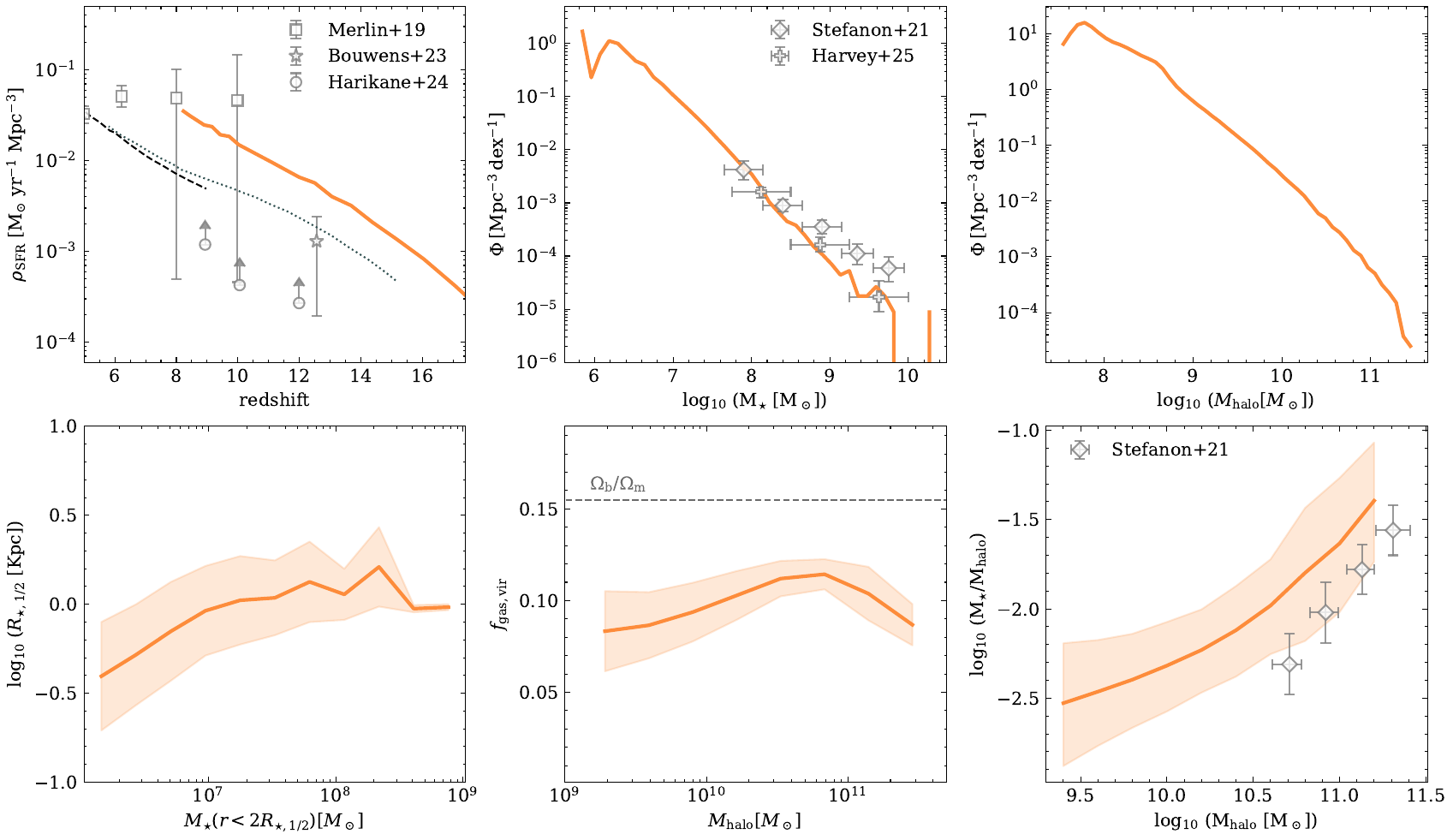}
    \caption{General properties of the \textsc{Phoebos} simulation. The top-left panel shows the temporal evolution of the cosmic SFRD, whereas all other panels show quantities at $z = 8$: the galaxy stellar mass function (top-centre panel), the total halo mass function (top-right panel), the sizes of galaxies as a function of stellar mass (bottom-left panel), the gas fraction of haloes as a function of the total halo mass, with the gray dashed line showing the cosmic baryon fraction \citep[][bottom-centre panel]{Planck_2018_cosmo_params}, and the stellar mass-halo mass relation (bottom-right panel). We compare the results from the \textsc{Phoebos} simulation with the observed cosmic SFRD from \citet{Merlin:2019aa, Bouwens:2023aa, Harikane:2024aa} and simulated data from \textsc{TNG100} \citep[][black dashed line]{Jo:2023aa} and \textsc{Thesan}-1 \citep[][gray dotted line]{Kannan:2022aa}. Furthermore, we have plotted the observed stellar mass functions (both at $z \sim 8$) from \citet{Stefanon:2021aa} and \citet{Harvey:2025aa} and the stellar mass-halo mass relation from \citet{Stefanon:2021aa}.}
    \label{fig:simprops}
\end{figure*}

The simulations were run with the $N$-body smoothed-particle hydrodynamics code \textsc{ChaNGa}, using a Wendland C4 kernel \citep[][]{Wendland_1995,Dehnen_Aly_2012} and 50 neighbours. We adopted both thermal (with a coefficient of 0.03) and metal (0.05) diffusion, as described in \citet{Shen_et_al_2010}. The gas is allowed to cool via several processes linked to primordial and metal species: primordial species and cooling are computed in non-equilibrium \citep[with self-shielding;][]{Pontzen_et_al_2008} assuming a redshift-dependent ultraviolet radiation background \citep[][]{Haardt_Madau_2012}, whereas metal cooling is calculated in photo-ionization equilibrium with the same background \citep[][]{Shen_et_al_2010,Shen_et_al_2013}, using tabulated rates from \textsc{Cloudy} \citep[][]{Ferland_et_al_1998,Ferland_et_al_2013}, thus assuming no self-shielding \citep[for a discussion, see][]{Capelo_et_al_2018}. We did not impose any pressure floor but assumed a cooling temperature floor\footnote{One exception is the Phoebos run, wherein the floor was set equal to 300~K up to $z \sim 10$.} of 10~K.

Stars form stochastically whenever the gas is colder than $3 \times 10^4$~K and denser\footnote{Because of the lower resolution, the density threshold in the \textsc{PhoebosLR} and \textsc{PhoebosULR} simulations was set to $0.1 \,m_{\rm p}$~g~cm$^{-3}$.} than $1.0 \,m_{\rm p}$~cm$^{-3}$, where $m_{\rm p}$ is the proton mass, according to the probability function $p = (m_{\rm gas}/m_{\rm star})[1 - \exp{(-\epsilon_{\rm SF} \Delta t/t_{\rm dyn})}]$, where $m_{\rm gas}$ and $m_{\rm star}$ are the masses of the (potentially) star-forming gas particle and new stellar particle (set equal to each other in our simulations), respectively, $\Delta t = 10^6$~yr is how often SF is computed, $t_{\rm dyn} = (4 \pi G \rho_{\rm gas})^{-1/2}$ is the local dynamical time, where $G$ is the gravitational constant, and $\epsilon_{\rm SF}$ is the SF efficiency (\citealt{Stinson_et_al_2006}; see also \citealt{Katz_1992}). This implies that, on average, ${\rm d} M_{\star}/{\rm d}t = \epsilon_{\rm SF} M_{\rm gas}/t_{\rm dyn}$, where $M_{\star}$ and $M_{\rm gas}$ are the mass of stars and available gas involved, effectively ensuring that SF approximately follows the slope of the Schmidt-Kennicutt relation between SFR and gas surface density \citep[][]{Schmidt_1959,Schmidt_1963,Kennicutt_1989, Kennicutt_1998}. The normalization of this relation is set by $\epsilon_{\rm SF}$, which is equal to 0.1 in all our simulations.

The newly formed stellar particles are much more massive than individual stars, their mass being $\sim$$10^6$~M$_{\sun}$. For this reason, they represent an entire stellar population, whose mass distribution is given by the \citet{Kroupa_2001} initial mass function. The individual stars of the population have different lifetimes, which depend on their mass and are computed according to \citet{Raiteri_et_al_1996}. At the end of their lifetime, stars with masses in the range 8--40~M$_{\sun}$ can explode as Type~II SNae, which inject into the surrounding ISM gas mass, iron mass, and oxygen mass, dependent on the mass of the progenitor star according to \citet{Raiteri_et_al_1996}, and $10^{51}$~erg of thermal energy, according to the `blastwave model' of \citet{Stinson_et_al_2006}, in which radiative cooling is disabled during the hot, low-density SN shell's survival time \citep[][]{McKee_Ostriker_1977}. Thermal energy ($10^{51}$~erg) and mass and metals (independent of progenitor stellar mass; \citealt{Thielemann_et_al_1986}; and with no radiative cooling disabled) are also injected in the surrounding gas from Type~Ia SNae, modelled to occur in binary systems of total mass ranging from 3 to 16~M$_{\sun}$, using the binary fractions from \citet{Raiteri_et_al_1996}. Finally, stars with masses in the range 1--8~M$_{\sun}$ release part of their mass (and metals) as stellar winds, using the recipes of \citet{Kennicutt_et_al_1994} and \citet{Weidemann_1987}.

\section{Results}\label{sec:results}

We present the initial findings from the \textsc{Phoebos} simulation, focusing on the evolution of galaxy properties in the early Universe. Figure~\ref{fig:simprops} shows the general properties of the \textsc{Phoebos} simulation at $z=8$ and, where possible, a comparison with observations or other simulations. The figure highlights key relationships between SFR, stellar mass, and total halo mass, providing insight into how galaxies evolve in the high-redshift Universe. The top-left panel shows the temporal evolution of the SFR density (SFRD) up to $z = 8$, compared to observational constraints  from \citet{Merlin:2019aa, Bouwens:2023aa} and lower limits from \citet{Harikane:2024aa}. The \textsc{Phoebos} simulation aligns well with the observations from \citet{Merlin:2019aa} and are above the lower limits of \citet{Harikane:2024aa}, but appears to slightly overestimate the SFRD relative to the data point from \citet{Bouwens:2023aa}. The SFRD of the \textsc{Phoebos} simulation is higher between $z \sim 8$ and 15 than other simulations of a similar box size, e.g. \textsc{TNG100} \citep[][]{Jo:2023aa} and \textsc{Thesan} \citep[][]{Kannan:2022aa}. The elevated SFRD in this redshift range suggests that \textsc{Phoebos} may capture different physical processes or feedback mechanisms compared to these other simulations. This makes \textsc{Phoebos} a key project for exploring the effects of a higher SFRD than previously tested in simulations, shaping galaxy properties such as stellar mass growth, metallicity, and ionizing photon production, and providing new insights into early galaxy formation and reionization.

\begin{figure}
    \centering
    \includegraphics[ trim={0cm 0cm 0cm 0cm}, clip, width=0.48\textwidth, keepaspectratio]{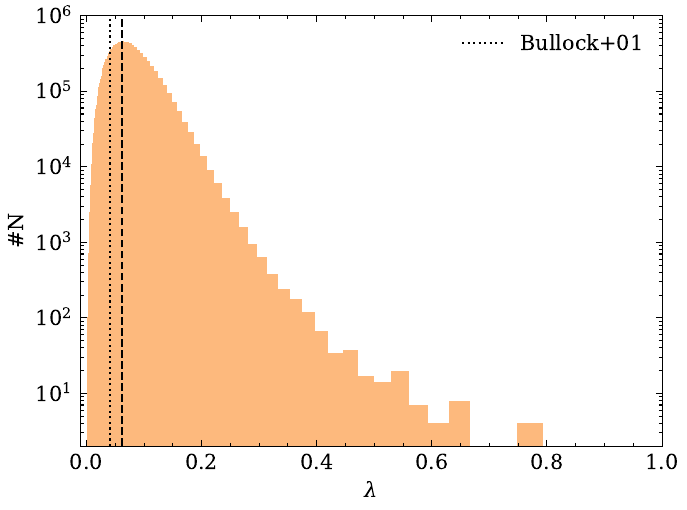}
    \caption{Spin parameter distribution of our simulated galaxies at $z=8$. The black dashed line indicates the peak of the distribution, which is consistent with the expected value from \citet{Bullock:2001aa}, shown for comparison as the black dotted line.}
    \label{fig:spin} 
\end{figure}

\begin{figure}
    \centering
    \includegraphics[ trim={0cm 0cm 0cm 0cm}, clip, width=0.48\textwidth, keepaspectratio]{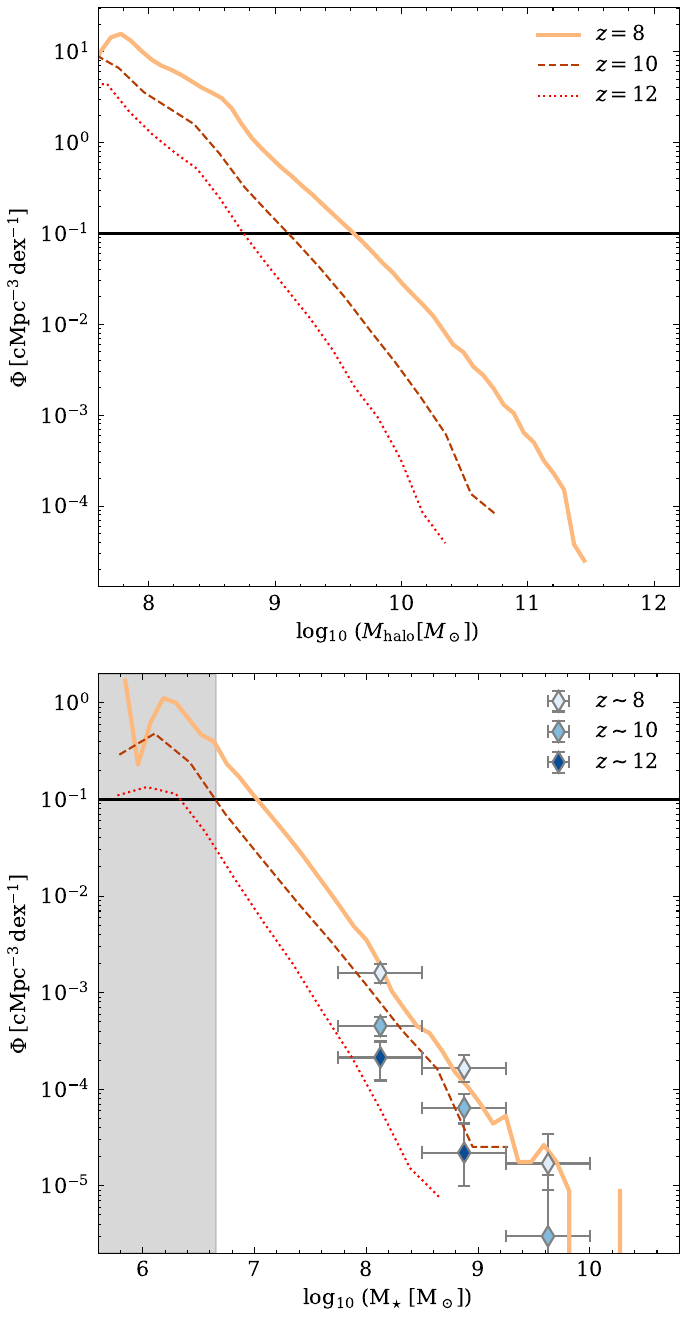}
    \caption{Total halo mass (top panel) and stellar mass (bottom panel) functions of the \textsc{Phoebos} simulation across different redshifts: $z = 8$ (orange solid line), $10$ (brown dashed line), and $12$ (red dotted line). The bottom panel also shows the observed stellar mass function from \citet{Harvey:2025aa}, with redshift bins $7.5 < z \leq 8.5$ (light blue), $9.5 < z \leq 11.5$ (blue), and $11.5 < z \leq 13.5$ (dark blue). The gray shaded region in the bottom panel highlights the mass range where haloes contain fewer than 5 stellar particles. The vertical black line marks the volume limit, corresponding to the “10 galaxies per dex” threshold.}
    \label{fig:massfunc} 
\end{figure}

The top-centre and top-right panels show the distributions of stellar mass and total (including baryons and DM) halo mass, respectively, at $z = 8$. We find $10^{7.05}$ haloes, identified using the adaptive-mesh \textsc{AMIGA Halo Finder} \citep[\textsc{AHF};][]{Gill:2004aa, Knollmann:2009aa} code. The masses are measured within the virial radius ($R_{\rm vir}$), defined using the overdensity threshold, $\Delta_{\rm vir}$, from \citet{Gross_1997} for a flat $\Lambda$CDM Universe. The stellar mass distribution is compared with observational data from \citet{Stefanon:2021aa} and \citet{Harvey:2025aa}. The \citet{Harvey:2025aa} data corresponds to an average redshift of 8.02, whereas the \citet{Stefanon:2021aa} data represents galaxies at approximately $z \sim 8$. Our simulation aligns very well with these observations. A more detailed discussion of the mass distributions and their evolution can be found in Section~\ref{sec:dist_highz}.

The bottom-left panel examines the relationship at $z = 8$ between galaxy stellar mass (defined as the enclosed stellar mass within twice the stellar half-mass radius) and stellar half-mass radius ($R_{\star, 1/2}$), a key indicator of galaxy size evolution. The shaded region represents the 1$\sigma$ scatter within each mass bin. The \textsc{Phoebos} simulation shows a correlation between these two quantities, with larger galaxies tending to be more massive, as expected. A more detailed discussion of the stellar size-to-mass relation can be found in Section~\ref{sec:dist_highz}.

Lastly, the bottom-centre panel presents the gas fraction within haloes, and the bottom-right panel presents the stellar-to-halo mass ratio (SHMR), both as a function of total halo mass, alongside observational data from \citet{Stefanon:2021aa}, all at $z = 8$. Our simulation aligns well with these observations, staying within the standard deviation. A more detailed discussion of the SHMR can be found in Section~\ref{sec:SHMR}.

To provide an additional check if the galaxies of our simulation are physical, Figure~\ref{fig:spin} presents the distribution of the spin parameter, computed by \textsc{AHF} at $z = 8$, with the peak of the distribution marked by a black dashed line. The distribution behaves as expected, and the peak value of $0.06$ aligns closely with the value found by \citet{Bullock:2001aa}, $0.04$, shown by the black dotted line, supporting the physical realism of our simulated galaxies.

\subsection{Mass and size distribution evolution}\label{sec:dist_highz}

Figure~\ref{fig:massfunc} presents the total halo mass and stellar mass functions (within $R_{\rm vir}$) computed by \textsc{AHF}, at redshifts $z = 8$, $10$, and $12$, illustrating the evolution of structure formation over cosmic time.  As expected, both mass functions exhibit a clear redshift dependence, with number densities decreasing at earlier epochs, reflecting the hierarchical build-up of structure. 

At $z = 12$, the number density of haloes is significantly lower compared to $z =8$ across all mass bins, particularly at the high-mass end (M$_{\rm halo} > 10^{10}$ M$_{\sun}$), where haloes remain rare. By $z = 10$, the overall normalization of the mass function increases, suggesting continued halo growth via accretion and mergers. At $z = 8$, this trend becomes more pronounced, with a substantial rise in the abundance of haloes, particularly in the intermediate- and high-mass ranges. This evolution is consistent with expectations from hierarchical structure formation, wherein small haloes form first and subsequently grow through mergers and accretion.

\begin{figure}
    \centering
    \includegraphics[ trim={0cm 0cm 0cm 0cm}, clip, width=0.48\textwidth, keepaspectratio]{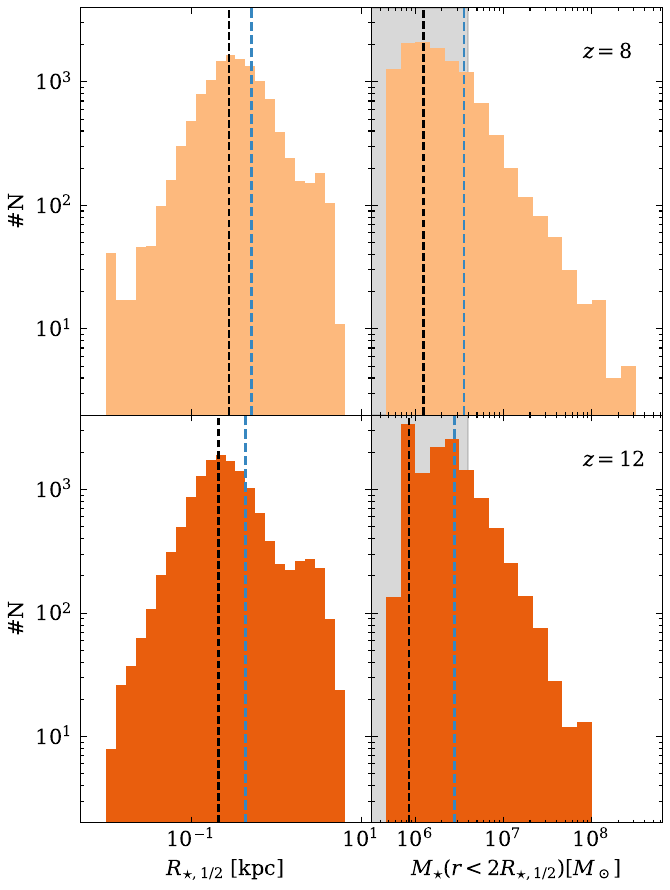}
    \caption{Evolution of stellar properties from $z=12$ to $z=8$. The top panels correspond to $z=8$, whereas the bottom panels represent $z=12$. The left-hand panels show histograms of the stellar half-mass radius. The right-hand panels present the stellar mass distributions. The blue dashed lines indicate the means of the distributions, whereas the black dashed line mark the peaks. The gray areas in the right-hand panels highlight the regions where haloes consist of fewer than 5 stellar particles: these are excluded in the following figures.}
    \label{fig:histRM}
\end{figure}

The stellar mass function (bottom panel) aligns well with the data from \citet{Harvey:2025aa}. The observed points in the $z = 7.5$--8.5 bin match the $z = 8$ simulation results, indicating good agreement with model predictions. At higher redshifts, some deviations emerge, the observed number density is higher than we find within the \textsc{Phoebos} simulation, especially at $z \sim 12$. Nevertheless, the simulated data largely remain within their error margins. 

\begin{figure}
    \centering
    \includegraphics[ trim={0cm 0cm 0cm 0cm}, clip, width=0.48\textwidth, keepaspectratio]{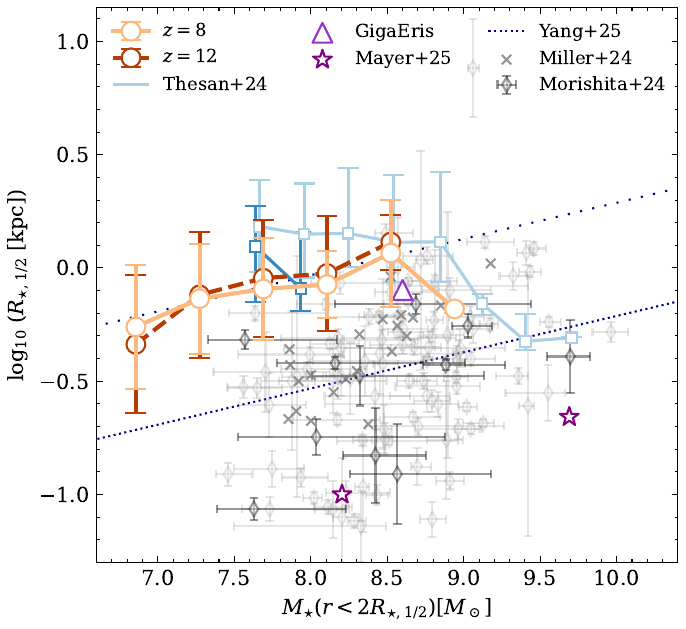}
    \caption{Stellar size–mass relation at $z =8$ and $12$, defined between the stellar half-mass radius and the stellar mass enclosed within twice that radius. The open circles and error bars represent the median and $1\sigma$ scatter of halo sizes from \textsc{Phoebos}. These results are compared to observational constraints on the effective radius from \citet{Morishita:2024aa} at two different redshift ranges, depicted as diamond markers, and \citet{Miller:2024aa} depicted as the gray crosses at $z=8$. Light gray diamonds correspond to observations in the redshift range $7.5 \lesssim z \lesssim 8.5$, whereas dark gray diamonds represent data in the range $11.5 \lesssim z \lesssim 12.5$. The dark blue dotted line represents the size–mass relation at $z \sim 7.5$ reported by \citet{yang:2025aa}, and the loosely dotted line shows the same relation with a modified $y$-intercept to better align with the simulations. Simulated results from the \textsc{Thesan} simulation \citep[][]{Shen:2024aa} are shown as blue square markers, with light blue representing $z \simeq 8$ and dark blue corresponding to $z \simeq 12$. Additionally, the two companion galaxies from the \textsc{MassiveBlackPS} simulation analysed in \citet{Mayer:2025aa} at $z \sim 7.6$ are shown as purple stars, and the main galaxy from the \textsc{GigaEris} simulation \citep{Tamfal:2022aa, vanDonkelaar:2023aa, Donkelaar:2025aa} at $z \sim 8$ is shown as a purple triangle. }
    \label{fig:RMobs}
\end{figure}

A similar trend is observed in the stellar mass function, which also exhibits a strong redshift dependence, also shown in Figure~\ref{fig:histRM} where we focus on the stellar mass within twice the stellar half-mass radius. At $z = 12$, the abundance of galaxies with significant stellar masses remains low, indicating that SF is still in its early stages. By $z = 10$, the number density increases and at $z = 8$ this trend continues, with a notable increase in the abundance of galaxies, particularly at the higher-mass end (M$_{\star} > 10^{9}$ M$_{\sun}$). This trend is also evident in the right-hand panels of Figure~\ref{fig:histRM}, where the mean stellar mass within twice the stellar half-mass radius of haloes increases from $10^{5.9}$ to $10^{6.1}$~M$_{\sun}$, and the maximum stellar mass from $10^{8.4}$ to $10^{9.1}$~M$_{\sun}$, between $z = 12$ and $z = 8$ (i.e. in $\sim$0.3~Gyr). The rapid evolution of the stellar mass function between $z = 12$ and $z = 8$ suggests efficient SF and mass growth over this period, as also shown in the top-left panel of Figure~\ref{fig:simprops}.

In the left-hand panels of Figure~\ref{fig:histRM}, one can also see the evolution of the stellar half-mass radius ($R_{\star, 1/2}$) from $z = 12$ to $z = 8$. At the lower redshift, the mean size of the galaxy has increased by a factor of 1.21 from the value at the higher redshift, reflecting significant growth over time. Nevertheless, by $z=8$, a greater number of smaller structures have formed. These smaller structures are substructures embedded within the larger haloes, suggesting a more complex and hierarchical formation process as the system evolves. From this point onward, to ensure a high enough particle resolution,  all haloes considered must contain a minimum of 100 baryonic particles (i.e. at least $\sim$$10^8$~M$_{\sun}$ of baryonic mass). Additionally, when discussing stellar properties, the criterion is set to include at least 5 stellar particles (i.e. at least $\sim$$ 4 \times 10^6$~M$_{\sun}$ of stellar mass). 

To continue looking at the sizes of these haloes, Figure~\ref{fig:RMobs} shows the stellar size–mass relation at $z = 8$ and $z = 12$, with a comparison between simulation results and observational data from \citet{Morishita:2024aa} at $7.5 \lesssim z \lesssim 8.5$ (light gray), and $11.5 \lesssim z \lesssim 12.5$ (dark gray). Furthermore, the dotted line represents the size–mass relation at $z \sim 7.5$ reported by \citet{yang:2025aa}. Notably, the observed galaxies appear to be consistently more compact than the predictions from the simulations, echoing the findings from the \textsc{Thesan} simulation, also shown in the figure \citep[light and dark blue squares;][]{Shen:2024aa}. This discrepancy is likely due to the difference between how sizes are defined in the theoretical models and the observed galaxies. While \citet{Morishita:2024aa} and \citet{yang:2025aa} report the effective radius, defined as the radius that encloses half of the galaxy's light, we and the work down by \citet{Shen:2024aa} are comparing it with the stellar half-mass radius for both simulations. At $z \gtrsim 8$, galaxies are typically quite compact and as a result, the effective radius would be smaller than the stellar half mass radius, which encloses half of the stellar mass\footnote{For a more detailed discussion see e.g. \citet{Kravtsov:2013aa} and \citet{Szomoru:2013aa}, the half-light radius of galaxies is offset by approximately 25 per cent from the half-mass radius $r_{1/2}$, regardless of galaxy stellar mass, morphology, or redshift.}. Therefore, the radii within \textsc{Phoebos} should be seen considered as an upper limit compared to the effective radius.

\begin{figure}
    \centering
    \includegraphics[ trim={0cm 0cm 0cm 0cm}, clip, width=0.47\textwidth, keepaspectratio]{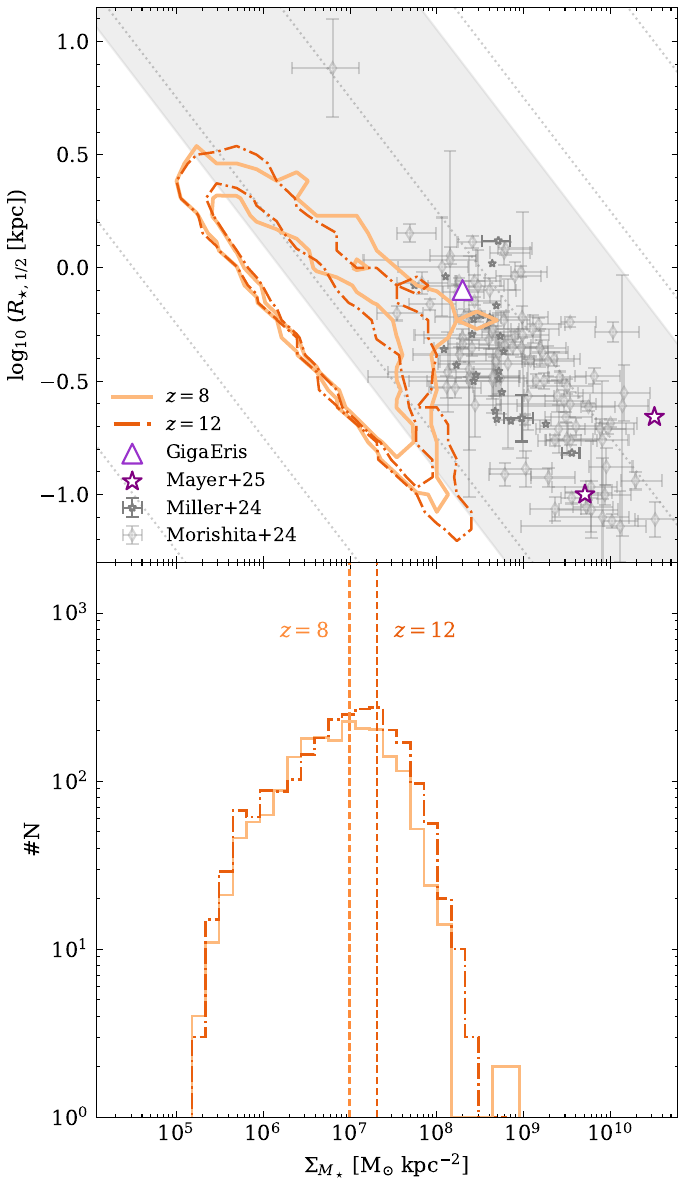}
    \caption{Stellar density evolution of the stellar component within haloes from $z = 12$ to $z = 8$. The top panel shows the correlation between the stellar half-mass radius and the stellar surface density, $\Sigma_{M_\star}$. Simulation results are compared with observational constraints from \citet{Miller:2024aa} at $z=8$ and \citet{Morishita:2024aa} in the range $7.5 \lesssim z \lesssim 8.5$. Gray dotted lines represent stellar haloes of similar mass assuming spherical symmetry, illustrating where observations would fall with different radius definitions. The gray-shaded region indicates where observed haloes would lie depending on radius. The bottom panel shows the stellar density distribution at the two redshifts. Dashed lines mark the peak stellar density: $10^{6.99}$~M$_{\sun}$~kpc$^{-2}$ at $z=8$ and $10^{7.31}$~M$_{\sun}$~kpc$^{-2}$ at $z=12$. The companion galaxies analyzed in \citet{Mayer:2025aa} at $z \sim 7.6$ are indicated as a purple stars, and the \textsc{GigaEris} galaxy \citep{Tamfal:2022aa, vanDonkelaar:2023aa, Donkelaar:2025aa} at $z \sim 8$ is shown as a purple triangle.}
    \label{fig:stellardens}
\end{figure}

\begin{figure*}
    \centering
    \includegraphics[ trim={0cm 0cm 0cm 0cm}, clip, width=1\textwidth, keepaspectratio]{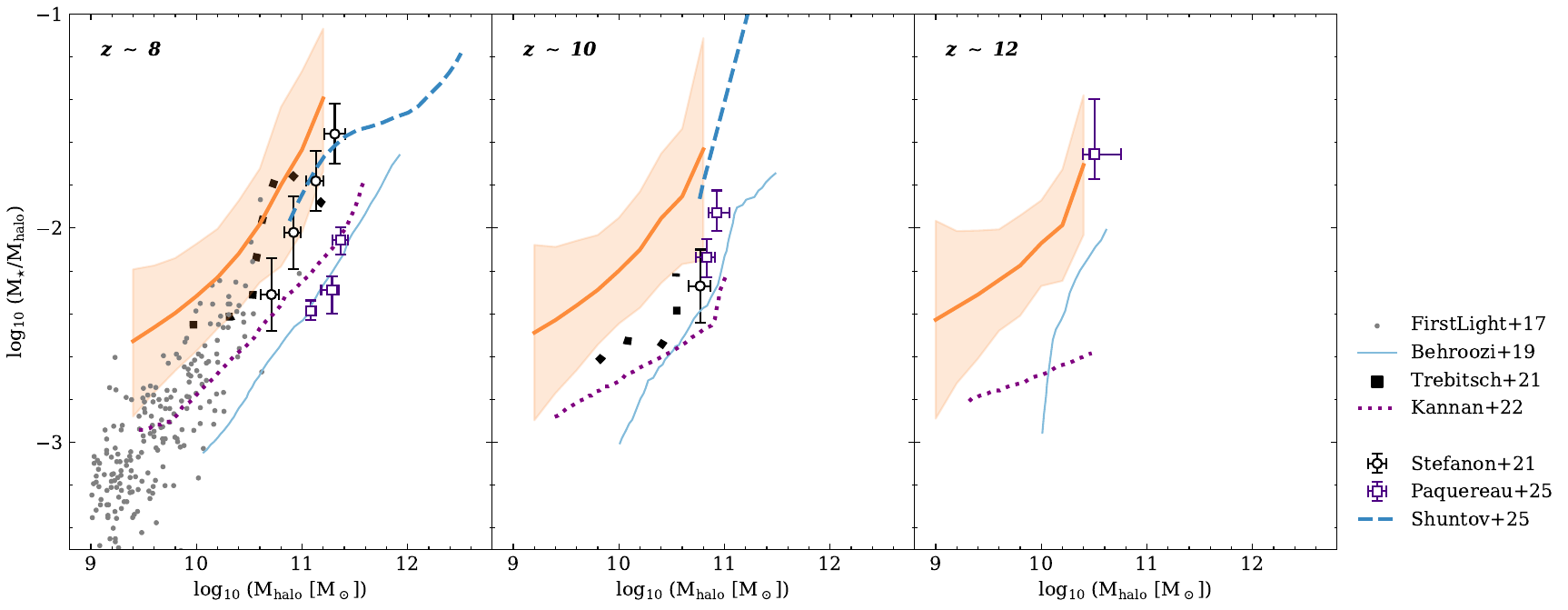}
    \caption{The stellar-to-halo mass relation (SHMR; orange solid line) and associated 1$\sigma$ scatter (orange-shaded region) for galaxies in the \textsc{Phoebos} simulation at $z = 8$ (left-hand panel), $z = 10$ (middle panel), and $z = 12$ (right-hand panel). Observational constraints are overlaid for comparison: black markers from \citet{Stefanon:2021aa}, blue dashed line from \citet{Shuntov:2025aa}, and purple markers from \citet{Paquereau:2025aa}. Theoretical predictions are also included: the UniverseMachine model from \citeauthor{Behroozi:2019aa} (\citeyear{Behroozi:2019aa}, solid teal line), the FirstLight simulation \citep[][gray markers]{Ceverino:2017ab}, the \textsc{Thesan}-1 simulation \citep[][purple dotted line]{Kannan:2022aa}, and the \textsc{Obelisk} simulation \citep[][black square markers]{Trebitsch:2021aa}. The data of \textsc{Thesan}-1, \textsc{Obelisk}, \citet{Behroozi:2019aa}, and \citet{Paquereau:2025aa} are in three redshift bins: $6 \leq z \leq 8$, $8 \leq z \leq 10.5$, and $10.5 \leq z \leq 14$.}
    \label{fig:SHMR}
\end{figure*}

Additionally, it has been extensively studied and demonstrated in the literature that, when the smoothing length is significantly smaller than the softening length in a self-gravitating system (which is our case, since the smoothing length's minimum can be 5 per cent of the softening), gravitational collapse is either slowed down or partially suppressed \citep[e.g.][]{Bate:1997aa}. This suggests that our haloes may appear more puffed up due to resolution constraints. This is further supported when analysing the two companion galaxies analysed in \citet{Mayer:2025aa} at $z \sim 7.6$ (purple stars). With a gravitational softening of 2~pc in the \textsc{MassiveBlackPS} simulation \citep[][]{Mayer_et_al_2024}, we indeed see that the radii for similar massed objects become smaller. Importantly, \textsc{MassiveBlackPS} employs a stellar feedback model that is identical\footnote{While the stellar feedback model is formally identical, resolution-dependent star formation thresholds lead to differences in the density of gas affected by the blastwave feedback \citep[see e.g.][]{guedes:2011aa, Soko:2017aa}} to the one used in \textsc{Phoebos}, allowing for a meaningful comparison. The same outcome arises when studying the main object from the cosmological `zoom-in' simulation \textsc{GigaEris} \citep{Tamfal:2022aa, vanDonkelaar:2023aa, Donkelaar:2025aa}, which follows a galaxy that should be quite common within the \textsc{Phoebos} simulation in size and halo mass, but with much better baryonic mass resolution ($798$~M$_{\sun}$) and softening (43~pc): we again see the drop in size for a similar mass. Crucially, \textsc{GigaEris} also adopts a stellar feedback prescription that is identical\footnotemark[8] to that of \textsc{Phoebos}, but has a gravitational softening in-between those of \textsc{MassiveBlackPS} and \textsc{Phoebos}, further reinforcing the interpretation that the observed size differences are driven by resolution. Informed with this notion, we show the loosely dotted line, representing the size-mass relation from \citet{yang:2025aa} with a modified $y$-intercept to better match the simulations. Notably, this adjusted relation closely follows the \textsc{Phoebos} results up to a stellar mass of $\sim 10^{9.5}$~M${\sun}$, successfully reproducing the slope, unlike the \textsc{Thesan} simulation \citep{Shen:2024aa} at a similar redshift. Above a stellar mass of $ \sim 10^{9.5}$~M$_{\sun}$, the number of haloes in the next mass bin drops to just four, making the number density too small for a robust comparison.

We observe that the sizes of galaxies at $z = 12$ and $z = 8$ are similar (for a given mass), suggesting that more massive galaxies undergo less significant size evolution over this period. This trend is also apparent when comparing with observational data from \citet{Morishita:2024aa}, where the dark gray points ($11.5 \lesssim z \lesssim 12.5$) and light gray points ($7.5 \lesssim z \lesssim 8.5$) lie within the same region of the parameter space. Similary, the haloes at $z = 12$ within \textsc{Phoebos} also appear to follow the slope reported by \citet{yang:2025aa} at $z \sim 7.5$. 

To further investigate stellar densities and in the attempt to understand if higher-redshift galaxies exhibit higher stellar densities, the evolution of stellar haloes from $z = 12$ to $z = 8$ is shown in Figure~\ref{fig:stellardens}. The top panel illustrates the relationship between the stellar half-mass radius, $R_{\star, 1/2}$, and the stellar surface density, $\Sigma_{M_\star}$\footnote{We assume spherical symmetry and calculate the galaxy’s surface density from its measured $R_{\star, 1/2}$ and $M_{\star}(r < 2R_{\star, 1/2})$ as $\Sigma_{M} = M / (\pi R^2$)}.  The orange contour line shows the distribution of galaxies at $=8$, while the dark orange dashed contour lines represent the results at  $z= 12$. As expected from Figure~\ref{fig:RMobs}, these distributions are very similar. These simulation results are compared with observational constraints of \citet{Miller:2024aa} at $z=8$ and \citet{Morishita:2024aa} in the range $7.5 \lesssim z \lesssim 8.5$. The gray-shaded region indicates where these haloes would lie by changing their radius while keeping the stellar mass fixed, since the radii used for the haloes in \textsc{Phoebos} represent the upper limit. Again, for guidance on how the radius and stellar surface density of the galaxies would change with better resolution, the results from \citet{Mayer:2025aa} and the \textsc{GigaEris} simulation at $z \sim 7.6$ and $\sim 8$, respectively, are shown in purple. Both of these points are clearly located in the gray-shaded region, being more dense than the galaxies identified in \textsc{Phoebos}. Since these simulations share the same stellar feedback, this suggests that we could be underestimating the stellar density due to resolution limitations. On the other hand, the fact that galaxies with lower stellar surface densities within the \textsc{Phoebos} simulation are not represented in the observational data  may be due to observational limitations or selection effects. 

As redshift increases, the peak stellar density shifts slightly toward higher values, as illustrated in the bottom panel, where the dashed lines denote the peaks of the stellar surface density distributions. Nonetheless, this change remains minimal. At $z=8$, the peak value is $10^{6.99}$ M$_{\sun}$~kpc$^{-2}$, increasing to $10^{7.31}$ M$_{\sun}$~kpc$^{-2}$ by $z=12$. While this trend suggests that higher-redshift galaxies may exhibit slightly elevated stellar densities and more compact sizes, consistent with denser stellar haloes at earlier cosmic times, the overall variation is subtle and should be interpreted with caution.

\subsection{Stellar-to-halo mass ratio}\label{sec:SHMR}

\begin{figure*}
    \centering
    \includegraphics[ trim={0cm 0cm 0cm 0cm}, clip, width=0.99\textwidth, keepaspectratio]{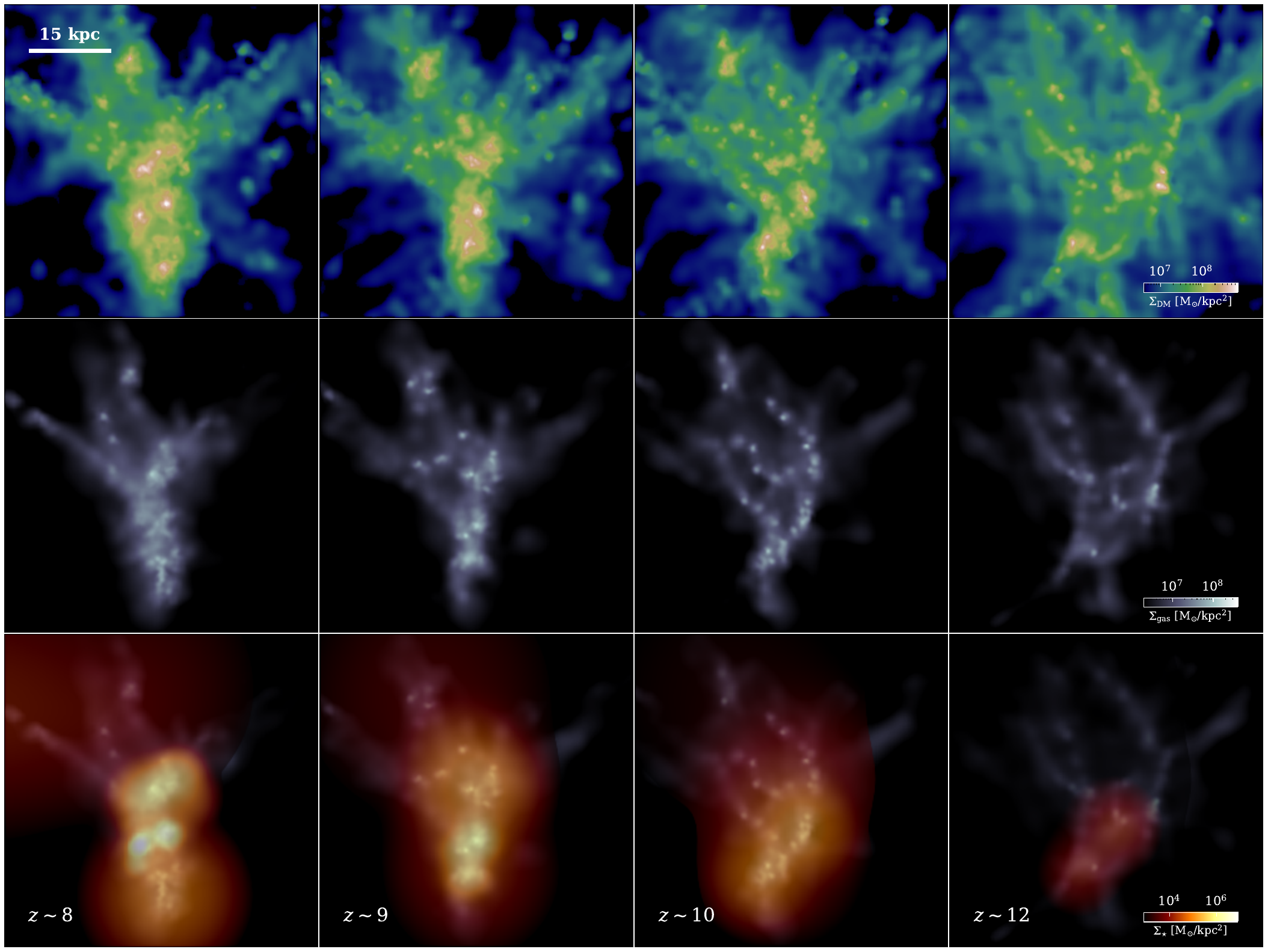}
    \caption{Surface density maps showing the evolution of a massive halo from $z \sim12$ and $z \sim 8$ in the \textsc{Phoebos} simulation. Rows show from top to bottom: DM surface density, gas surface density, and stellar surface density overlaid on the same gas map of the second row. The halo shown has a total mass of $10^{10.2}$~M$_{\sun}$, with a stellar mass of $10^{8.1}$~M$_{\sun}$ at $z \sim 8$. The virial radius at this redshift is 81.5~kpc. We show the halo within $0.7 \times R_{\rm vir, z = 8}$ (= 56.8 kpc) at all redshifts using the same physical volume.}
    \label{fig:galevo}
\end{figure*}

Figure~\ref{fig:SHMR} illustrates the SHMR, based on masses within the virial radius, for galaxies\footnote{Identified for haloes with zero substructures and a stellar mass above zero, giving us a total of $10^{5.9}$ galaxies.} in the \textsc{Phoebos} simulation at redshifts $z = 8$ (left-hand panel), $z = 10$ (middle panel), and $z = 12$ (right-hand panel).  The solid orange line represents the median relation, while the orange-shaded region denotes the 1$\sigma$ scatter across the simulation. For comparison, we include observational constraints from \citeauthor{Stefanon:2021aa} (\citeyear{Stefanon:2021aa}; black circular markers), \citeauthor{Shuntov:2025aa} (\citeyear{Shuntov:2025aa}; blue dashed line), and \citeauthor{Paquereau:2025aa} (\citeyear{Paquereau:2025aa}; squared purple markers). Furthermore, we overlay the theoretical predictions of \citeauthor{Behroozi:2019aa} (\citeyear{Behroozi:2019aa}; UniverseMachine, solid teal line) and results from the FirstLight simulation \citep[][gray markers]{Ceverino:2017ab}, \textsc{Thesan-1} \citep[][purple dotted line]{Kannan:2022aa}, and \textsc{Obelisk} \citep[][black square markers]{Trebitsch:2021aa}.

\begin{figure}
    \centering
    \includegraphics[ trim={0cm 0cm 0cm 0cm}, clip, width=0.32\textwidth, keepaspectratio]{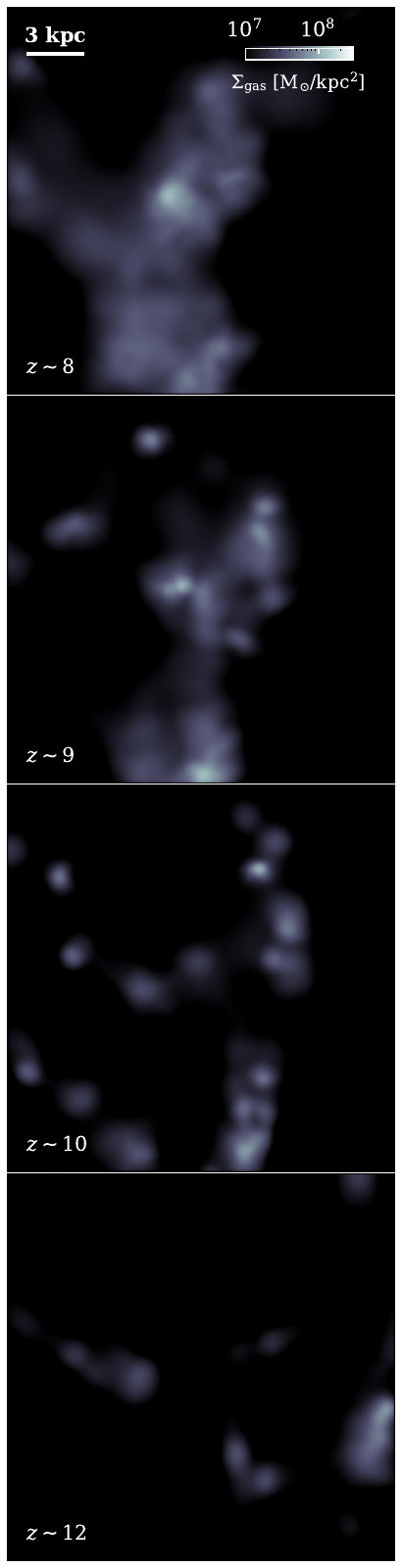}
    \caption{Zoom in of the gas surface density maps showing the evolution of the massive halo from $z \sim12$ and $z \sim 8$ of Figure~\ref{fig:galevo}.}
    \label{fig:galevoZoom}
\end{figure}

At $z \sim 8$ (left panel), the SHMR from the \textsc{Phoebos} simulation falls within the 1$\sigma$ scatter of the observational data from \citet{Stefanon:2021aa} and \citet{Shuntov:2025aa}, although it tends to overpredict stellar mass fractions relative to \citet{Paquereau:2025aa}. A similar offset with respect to \citet{Paquereau:2025aa} is also seen in the FirstLight and \textsc{Obelisk} simulations. Among these, \textsc{Phoebos} yields higher stellar mass fractions at fixed halo mass, especially when compared to FirstLight and \textsc{Thesan}-1, bringing it closer to the observational results of \citet{Shuntov:2025aa} at the high-mass end. 

Interestingly, while the SMHM relation from \citet{Stefanon:2021aa} falls below the median \textsc{Phoebos} curve, the stellar mass function is higher than the \textsc{Phoebos} results (see upper middle panel of Figure~\ref{fig:simprops}). This apparent inconsistency echoes the findings of \citet{Munshi:2013aa}, who showed that realistic feedback, baryon-induced halo mass suppression, and photometric stellar mass estimates can all reduce the inferred SMHM while preserving agreement with the stellar mass function. Another possible explanation for this discrepancy is the limited simulation volume. The galaxy number density could be underestimated due to the absence of rare overdense regions, which are capable of hosting a significantly larger population of galaxies at early epochs.

At higher redshifts ($z \sim 10$ and $z \sim 12$), the SHMR follows a similar trend but with systematically lower stellar mass fractions, consistent with the delayed SF in early haloes. Interestingly, the agreement between \textsc{Phoebos} and observational constraints appears to improve with increasing redshift. By $z \sim 12$ (right-hand panel), \textsc{Phoebos} is the only model that closely matches the trend reported by \citet{Paquereau:2025aa}. However, current observational constraints and simulated data at these redshifts remain minimal. Nonetheless, the results suggest that \textsc{Phoebos} may offer a promising match to the SHMR in the early Universe. 

Simulations such as TNG100 \citep[e.g.][]{Nelson_et_al_2018, Nelson:2019aa, Pillepich:2018aa, Marinacci:2018aa, Naiman:2018aa} and Horizon-AGN \citep[][]{Dubois:2014aa, kaviraj:2017aa}, as well as more recent efforts like \textsc{Thesan-Zoom} \citep{Kannan:2025aa}, have been remarkably successful at reproducing the SHMR at low redshifts ($z \lesssim 3$), probably due to their implementation of strong stellar feedback models. However, these simulations tend to exhibit minimal evolution at higher redshifts, particularly failing to capture the slope changes observed in recent JWST data \citep[see, e.g.][]{Paquereau:2025aa}. This suggests that their feedback prescriptions may not adequately account for the changing physical conditions of early galaxy formation. In this context, \textsc{Phoebos}, with its weak stellar feedback, stands out as one of the only current simulation that successfully reproduces the full high-redshift SHMR ($z \gtrsim 8$), giving us our first hints for the need of a redshift-dependent feedback regime in shaping early galaxy formation. 

\begin{figure}
    \centering
    \includegraphics[ trim={0cm 0cm 0cm 0cm}, clip, width=0.475\textwidth, keepaspectratio]{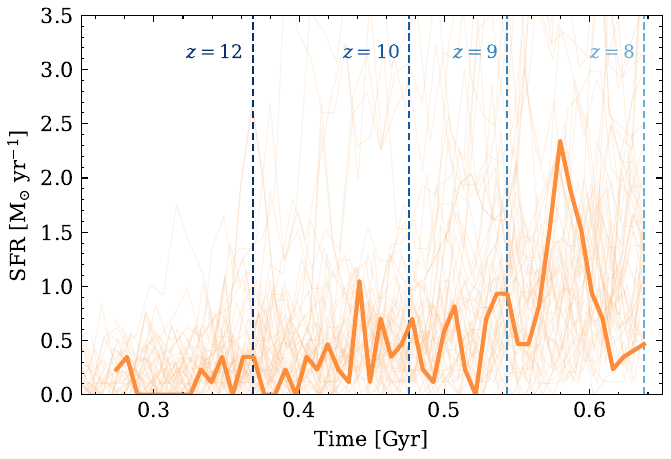}
    \caption{The SFR within the virial radius of the massive halo depicted in Figures~\ref{fig:galevo} and \ref{fig:galevoZoom}. The vertical dashed lines mark the redshifts corresponding to the snapshots presented in the surface density maps of the same figures. For comparison, light orange lines in the background represent the SFR histories of the 50 most massive haloes.}
    \label{fig:SFRmassive}
\end{figure}

\subsection{Halo formation}

\begin{figure*}
    \centering
    \includegraphics[ trim={0cm 0cm 0cm 0cm}, clip, width=1\textwidth, keepaspectratio]{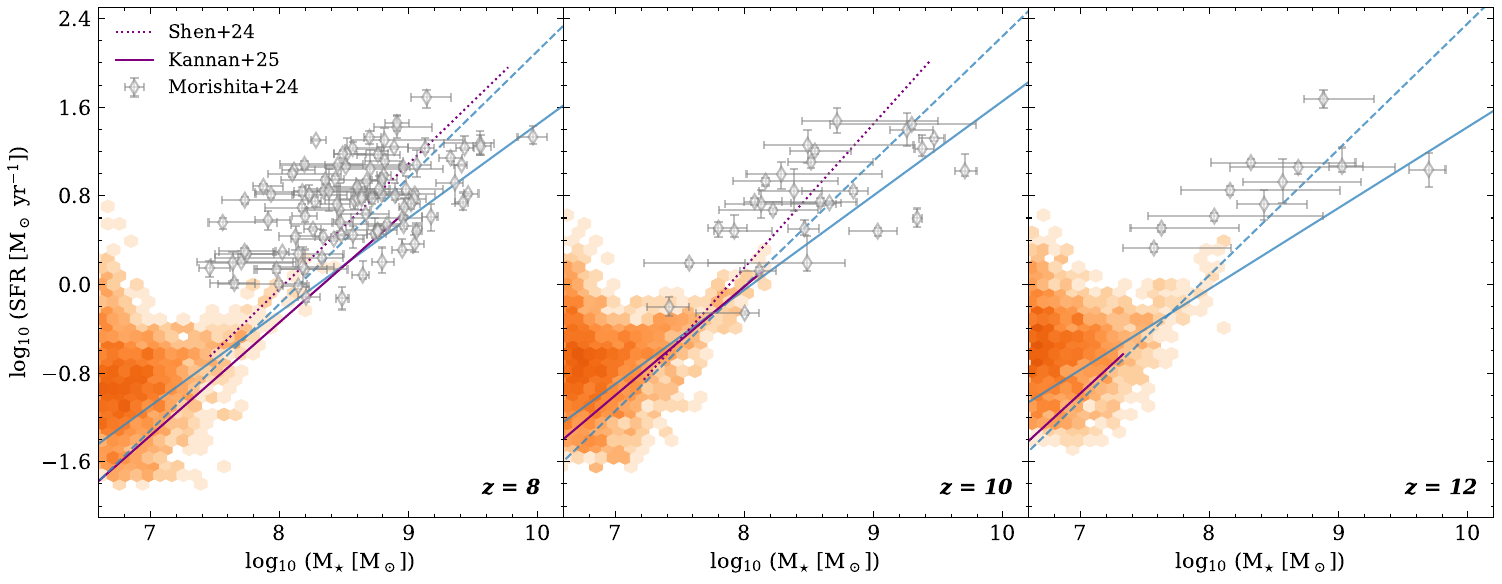}
    \caption{The SF main sequence of galaxies in \textsc{Phoebos} at $z= 8$, $10$, and $12$, from left to right, shown by the orange hexagrams. The gray diamonds correspond to observations in the redshift range $7.5 \lesssim z \lesssim 8.5$, $9.5 \lesssim z \lesssim 10.5$, and $11.5 \lesssim z \lesssim 12.5$ found by \citet{Morishita:2024aa}. The solid blue line represents the best-fit relation including the full simulated data, following the form  $\rm{SFR} \propto M_{\star}^{\alpha}$. The dashed blue line indicates the best-fit relation restricted to galaxies with stellar masses above the minimum observed by \citet{Morishita:2024aa} at each redshift. For comparison, the purple dotted line shows the relation from the \textsc{Thesan}-1 simulation \citep{Shen:2024aa}, while the solid purple line corresponds to results from the \textsc{Thesan-Zoom} simulations \citep{Kannan:2025aa}.}
    \label{fig:SFR}
\end{figure*}

Figure~\ref{fig:galevo} illustrates the evolution and formation of a massive galaxy halo from $z = 12$ to $z = 8$. The halo was identified at $z = 8$, where it ranks as the 25th most massive in terms of halo mass. The top row shows the DM surface density. Over time, the DM distribution becomes more concentrated and structured. By $z=8$, there are roughly four prominent overdensities within the halo, likely sites where the early galaxies are residing. The central row shows the corresponding gas surface density maps. The gas traces the filamentary structure of the DM but remains relatively low in density. As redshift decreases, the gas gradually condenses towards the halo centre, becoming increasingly clumpy and dense. This transition marks the onset of collapse processes that eventually lead to SF. This is captured in the bottom row, where the stellar surface density is overlaid on the gas surface density maps presented in the middle row, highlighting the relationship between stars and gas

For a closer look at the gas evolution, we zoomed into the centre of the halo in Figure~\ref{fig:galevoZoom}, where it is clear that the amount of gas structures decreases with increasing redshift. This is the result of ongoing halo mergers and the progressive assembly of mass towards the centre. Individual gaseous substructures, which are more prominent at higher redshifts, gradually blend and collapse into a more coherent, dense core.  This structural evolution is further supported by the SFR history shown in Figure~\ref{fig:SFRmassive}. The vertical dashed lines mark the redshifts corresponding to the DM, gas, and stellar surface density maps in Figures~\ref{fig:galevo} and~\ref{fig:galevoZoom}. We see that the SFR increases steadily as redshift decreases, peaking just before $z=8$, aligning with the visual progression of gas condensation and central clumping observed in the density maps. The sharp rise in SFR between $z\sim10$ and $z\sim8$ reflects the onset of efficient cooling and collapse processes, culminating in the formation of the stellar components highlighted in the bottom panels of Figure~\ref{fig:galevo}.

\subsubsection{Star formation}

\begin{figure}
    \centering
    \includegraphics[ trim={0cm 0cm 0cm 0cm}, clip, width=0.47\textwidth, keepaspectratio]{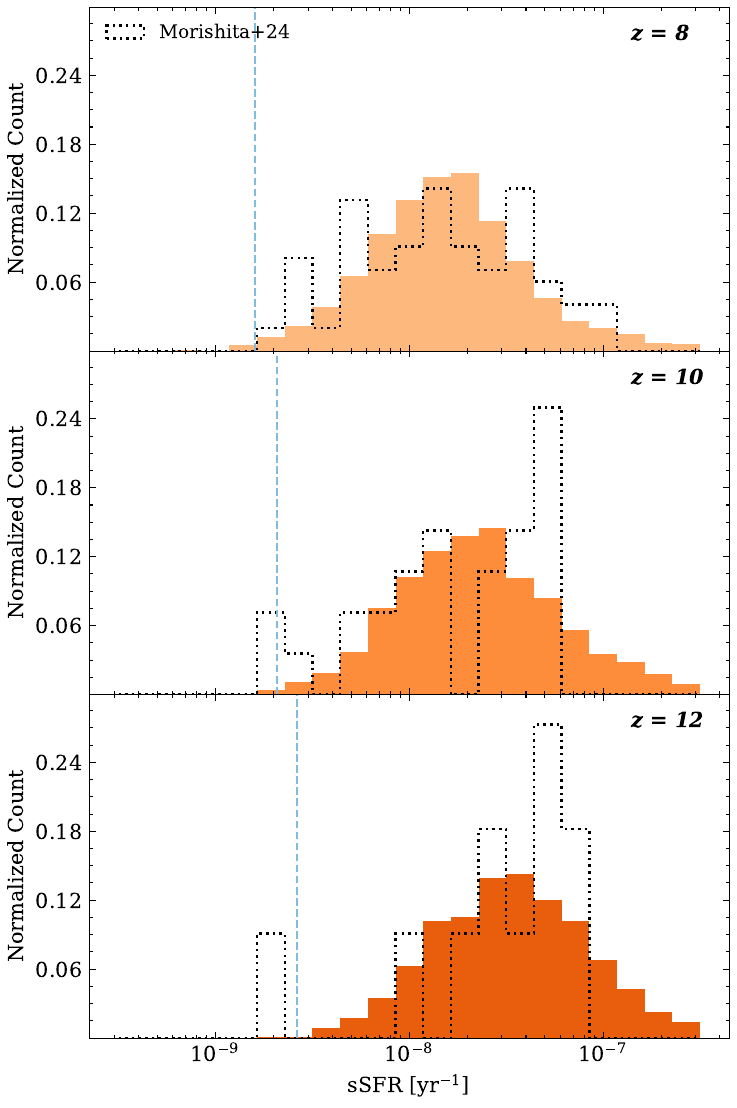}
    \caption{The normalized sSFR distribution at $z= 8$, $10$ and $12$ from top to bottom. The gray lines indicate the observed sSFR distribution in the redshift range $7.5 \lesssim z \lesssim 8.5$, $9.5 \lesssim z \lesssim 10.5$ and $11.5 \lesssim z \lesssim 12.5$ by \citet{Morishita:2024aa}. The blue vertical dashed line in each panel indicates $\rm{sSFR} = 1 / t_{\rm{univ}}(z)$, the specific star formation rate required to build up a galaxy’s stellar mass over the age of the universe at that redshift.}
    \label{fig:sSFR}
\end{figure}

Continuing our review of halo formation and galaxy properties, we now turn to the SF in the haloes.  Figure~\ref{fig:SFR} presents the average SFR of each galaxy over the past 50 Myr at $z= 8$, $10$, and $12$ from left to right, alongside the observations of \citet{Morishita:2024aa}. The Figure shows that the SFR within the \textsc{Phoebos} simulation closely follows the expected correlation between stellar mass and SFR found by \citet{Morishita:2024aa}. However, since our simulation is not constrained by observational lower limits, we detect more haloes at the lower-mass end compared to the observations. Conversely, our limited simulation box size prevents the formation of the most massive structures, which also explains the absence of the largest galaxies. This effect can similarly be observed in Figure~\ref{fig:RMobs}.

The solid blue line in Figure~\ref{fig:SFR} represents the best-fit relation through the simulated data, following the form  $\rm{SFR} \propto M_{\star}^{\alpha}$. The best-fit power-law indices are $\alpha = 0.85 \pm 0.03$, $0.85 \pm 0.01$, and $0.73 \pm 0.02$ for $z = 8$, $10$, and $12$, respectively. In contrast, the dashed blue line shows the best-fit relation when restricting to stellar masses above the minimum mass reported by \citet{Morishita:2024aa}. In this case, the slopes are steeper, with $\alpha = 1.14 \pm 0.1$, $1.12 \pm 0.02$, and $1.13 \pm 0.09$ at the same redshifts. Consistent with \citet{Morishita:2024aa}, we observe no clear declining trend in these indices, suggesting that SF efficiency remains high even at the high-mass end. This implies that, if the observed SF efficiency persists over the next $\sim$1~Gyr, some of our sources could evolve to stellar masses of $\sim$$10^{10}$~M$_{\sun}$ by $z \sim 4$, similar to the stellar mass of Milky Way-sized galaxies studied in detail in the \textsc{GigaEris} simulation \citep[][]{Tamfal:2022aa,vanDonkelaar:2023aa,Donkelaar:2025aa}. Nevertheless, \citet{Morishita:2024aa} report slightly higher SFRs compared to the solid blue line, meaning the normalization of the relation is off. This is likely caused by scatter in the SFRs of lower-mass galaxies, since the dashed blue line follows the observations more closely.

The positive correlation between SFR and halo mass is also consistent, with trends seen in other simulations.  For example, at both $z\sim10$ and $z\sim8$, the \textsc{Thesan}-1 simulation shows a clear increase in SFR with halo mass \citep{Yeh:2023aa} and a corresponding trend with stellar mass \citep{Shen:2024aa}. Furthermore, \citet{Shen:2024aa} report a very similar correlation between SFR and halo mass as seen in the blue dashed lines of the \textsc{Phoebos} simulation and observations, despite differences in stellar feedback models. Interestingly, the high-resolution \textsc{Thesan-Zoom} simulations (solid purple), which include a more sophisticated ISM model and implement an “early stellar feedback” mechanism compared to \textsc{Thesan}-1, recover an SFR–stellar mass relation with a slope similar to that reported by \citet{Shen:2024aa}. At lower stellar masses and especially at $=12$, \textsc{Thesan-Zoom} and \textsc{Phoebos} (blue dashed line) agree closely. Since both \textsc{Phoebos} and \textsc{Thesan-Zoom} employ this more advanced ISM model, despite differing feedback models and resolution, it suggests that the ISM model is crucial for achieving the SFR-stellar mass relation at $z \gtrsim 8$. However, resolution effects likely contribute to the scatter in SFR observed in \textsc{Phoebos}.

\begin{figure*}
    \centering
    \includegraphics[ trim={0cm 0cm 0cm 0cm}, clip, width=0.99\textwidth, keepaspectratio]{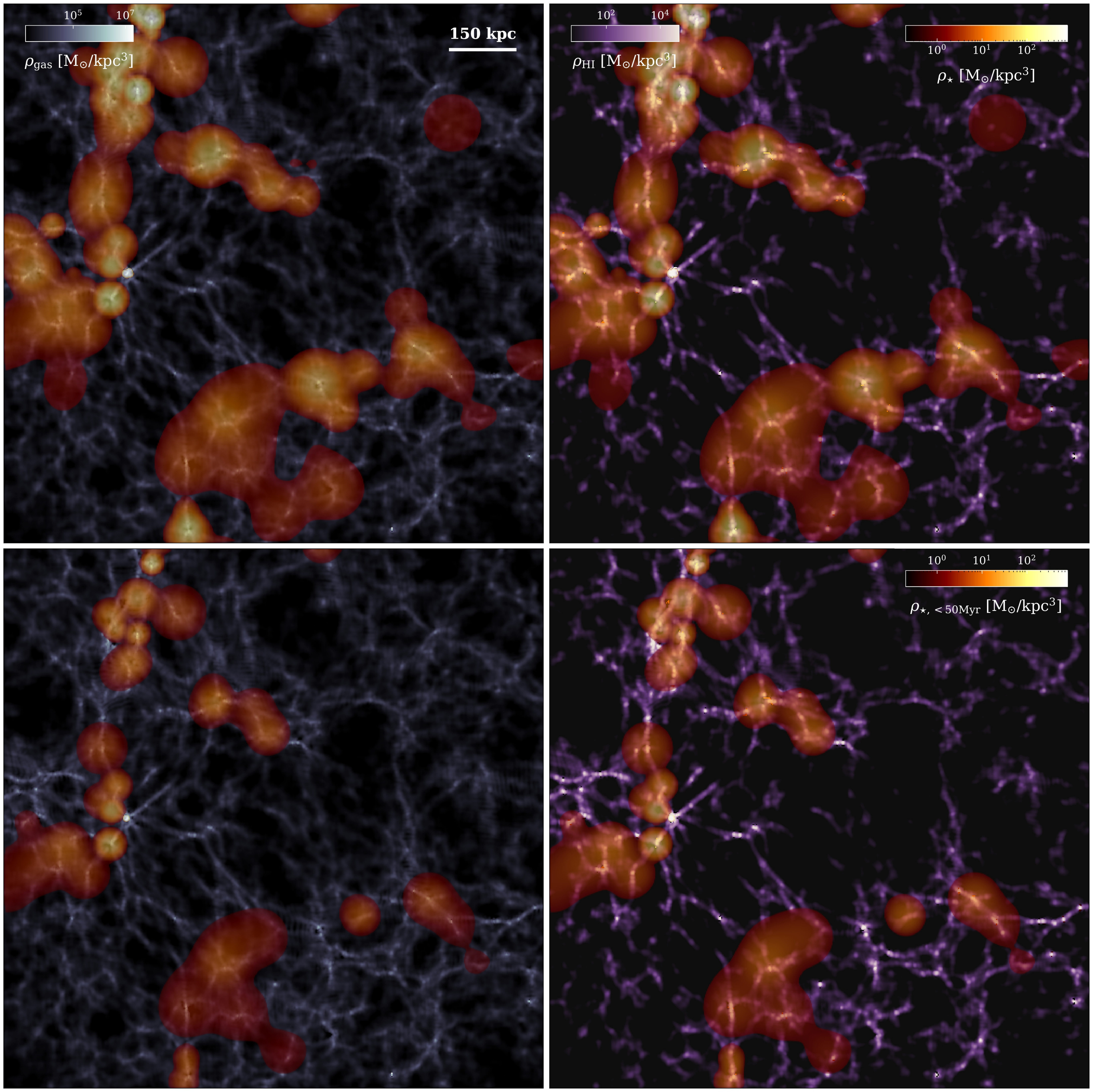}
    \caption{Gas volume density (left-hand panels) and HI volume density (right-hand panels) maps at $z=8$, shown alongside the stellar density distribution. The top panels display the full stellar population, while the bottom panels focus on the subset of young stars with ages $<50$~Myr. }
    \label{fig:HI_gas_star}
\end{figure*}

To further explore the idea of \textsc{Phoebos} having a realistic treatment of feedback and SF in this redshift range, we analyze the specific SFR (sSFR) distribution, shown in Figure~\ref{fig:sSFR} at $z= 8$, $10$, and $12$ together with observations of \citet{Morishita:2024aa}, which gives us valuable insight into the efficiency of SF relative to stellar mass. Figure~\ref{fig:sSFR} shows that, as we move to higher redshifts, the sSFR increases, suggesting that SF is more efficient at earlier epochs. This trend indicates that galaxies at higher redshift form stars at a faster rate relative to their stellar mass, in line with observations \citep[e.g.][]{Leslie:2020aa, Calaboro:2024}. Compared to observations, the \textsc{Phoebos} simulation once again shows excellent agreement with high-redshift data, reinforcing its ability to capture the key processes driving early galaxy formation. 

Furthermore, Figure~\ref{fig:sSFR} includes a blue vertical dashed line at $\rm{sSFR} = 1 / t_{\rm{univ}}(z)$, which represents the specific star formation rate needed for a galaxy to form its current stellar mass over the age of the Universe at that redshift. All galaxies in our sample lie above this line, indicating they are forming stars efficiently and rapidly, on timescales shorter than the Hubble time.

\subsection{Atomic neutral hydrogen density}

Atomic neutral hydrogen (HI) mapping plays a vital role in understanding galaxy evolution, as HI serves as a crucial gas reservoir for future SF, even though molecular gas is the immediate fuel for SF episodes \citep[e.g.][]{Tacconi:2020aa, Popping:2014aa, Wang:2020ab, Saintonge:2022aa, Gensior:2024aa}. HI is particularly sensitive to stellar feedback processes like SNae, which can stir turbulence and enhance velocity dispersion \citep[e.g.][]{Bacchini:2020aa, Silich:2001aa, Boomsma:2008aa, Orr:2022aa}. Moreover, HI content is closely tied to galaxy properties: galaxies with higher HI fractions tend to be bluer, more actively star-forming, and less metal-rich at fixed stellar mass, reflecting established correlations with stellar mass, morphology, and SF activity \citep[e.g.][]{Fumagalli:2008aa, Cortese:2011aa, Hughes:2013aa}.

The \textsc{Phoebos} simulation could offer a powerful platform for generating synthetic maps of neutral hydrogen (HI)in the Universe, enabling detailed studies of the large-scale structure traced by HI. These simulated maps provide a valuable framework for comparison with current and upcoming HI surveys, such as the Westerbork HI survey of SPiral and irregular galaxies \citep[WHISP;][]{Swaters:2002aa}, The HI Nearby Galaxy Survey \citep[THINGS;][]{Walter:2008aa}, the \textsc{bluedisk} project \citep{Wang:2013aa}, xGASS \citep{Catinella:2018aa}, and the pilot surveys of the Square Kilometre Array (SKA) precursors, such as the MeerKAT International GHz Tiered Extragalactic Exploration \citep[MIGHTEE;][]{Jarvis:2016aa} on MeerKAT and the Widefield ASKAP L-band Legacy All-sky Blind surveY \citep[WALLABY;][]{Koribalski:2020aa, Obeirne:2025aa} on the Australian SKA Pathfinder (ASKAP).

Figure~\ref{fig:HI_gas_star} demonstrates the potential of this approach by presenting the gas density (left-hand panels) and HI density (right-hand panels) alongside stellar density maps, for both the full stellar sample (top panels) and young stars ($<50$ Myr, bottom panels). From this figure, we can see that the HI gas fraction in the halo is indeed correlated with SF \citep[see also][]{Wang:2020ab, Lee:2025aa}. The regions with very high HI gas density also have higher stellar mass density, especially for the young stars, compared to the rest of the region plotted.

\begin{figure}
    \centering
    \includegraphics[ trim={0cm 0cm 0cm 0cm}, clip, width=0.48\textwidth, keepaspectratio]{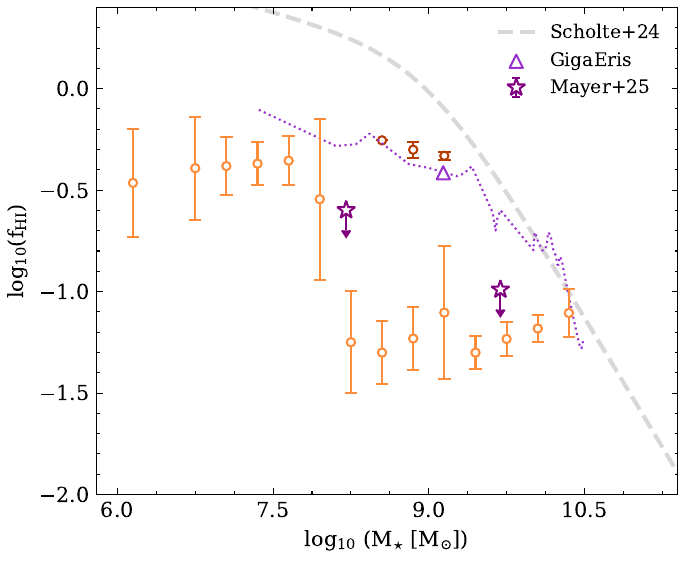}
    \caption{The stellar mass-HI gas fraction relation at $z = 8$. The open circles and error bars represent the median and $1\sigma$ scatter of the HI gas fraction of galaxies in \textsc{Phoebos}. These results are compared to observational data from \citet{Scholte:2024aa} for local galaxies with $0 < z < 0.6$, shown as the dashed gray line. The purple dotted line indicates the HI gas fraction evolution from the \textsc{GigaEris} simulation \citep{Tamfal:2022aa, vanDonkelaar:2023aa, Donkelaar:2025aa} up to $z \sim 4.4$, with a triangle marking $z \sim 8$. To enable a fairer comparison, dark orange open circles and error bars represent the HI gas fraction of a subset of \textsc{Phoebos} galaxies matched in halo and stellar mass both within 0.5 dex to the main galaxy of the \textsc{GigaEris} simulation. Additionally, the purple stars represent the two companion galaxies analyzed by \citet{Mayer:2025aa} at $z \sim 7.6$. }
    \label{fig:HI_gas}
\end{figure}

Figure~\ref{fig:HI_gas} shows the relationship between the HI gas fraction, f$_{\rm HI} = M_{\rm gas, HI}/ (M_\star + M_{\rm gas})$, and stellar mass for each galaxy in the sample, which includes only haloes without substructures and with $M_{\star} > 0$. In general, we find that more massive galaxies tend to have lower HI gas fractions, consistent with the trend identified in the local Universe by \citet{Scholte:2024aa} using data from the Dark Energy Spectroscopic Instrument. However, the HI gas fractions in our sample are significantly lower than those observed locally and exhibit an approximately flat trend for stellar masses above $M_{\star} > 10^8$~M$_{\sun}$. Additionally, we note a distinct drop in HI gas fraction around a stellar mass of $\sim$$10^8$~M$_{\sun}$ within the \textsc{Phoebos} simulation. While the exact origin of this feature remains uncertain, we will explore potential explanations in subsequent sections. It may point to a transition in galaxy evolution or a shift in the dominant feedback processes that become increasingly effective around this stellar mass scale.

We also compare our results to the \textsc{GigaEris} simulation \citep[purple dotted line for the full evolution up to $z \sim 4.4$, with a triangle marking $z \sim 8$,][]{Tamfal:2022aa, vanDonkelaar:2023aa, Donkelaar:2025aa} and the two companion galaxies studied in \citet{Mayer:2025aa} from the \textsc{MassiveBlackPS} simulation at $z \sim 7.6$ (purple stars). The \textsc{MassiveBlackPS} markers should be interpreted as upper limits, since the simulation does not include the HI fraction of the gas particles; instead, we display the total gas fraction. Both of these are zoom-in simulations with baryonic mass resolution smaller than $10^{3.5}$~M$_{\sun}$. Despite differences in redshift and simulation setups, both the \textsc{GigaEris} and \textsc{MassiveBlackPS} simulations exhibit trends similar to those observed in \textsc{Phoebos} and DESI, with the HI gas fraction decreasing as a function of stellar mass. However, \textsc{GigaEris} does not reproduce the sharp drop in HI gas fraction around $10^8$~M$_{\sun}$ seen in our sample, and the data point at $z \sim 8$ (purple triangle) shows a significantly higher HI gas fraction by about an order of magnitude compared to galaxies in that mass bin.

To enable a fairer comparison between \textsc{GigaEris} and \textsc{Phoebos}, the dark orange open circles and error bars represent the HI gas fraction of a subset of \textsc{Phoebos} galaxies matched in halo and stellar mass\footnote{See Appendix~\ref{appen:gig}, for a visual example of a \textsc{Phoebos} galaxy selected using this matching criteria.} both within 0.5 dex of the galaxy in the \textsc{GigaEris} simulation (purple triangle). These data points closely follow the purple dotted line representing \textsc{GigaEris} and lie near the triangle marker, indicating that this level of HI content is typical for galaxies with similar halo masses. The discrepancy between the full \textsc{Phoebos} sample and the \textsc{GigaEris}-like subset likely arises from the broader range of environments and halo masses sampled in \textsc{Phoebos}, which could contribute to the drop in HI content observed in the full sample. The \textsc{GigaEris}-like galaxies in \textsc{Phoebos} share similar halo mass scales with the galaxy in the \textsc{GigaEris} simulation, helping to explain their consistent gas fractions. Additionally, the large cosmological volume of \textsc{Phoebos} includes many lower-mass haloes where HI is more easily depleted. \textsc{GigaEris}, being a high-resolution zoom-in simulation, may also better resolve the internal gas dynamics that support the retention of HI.

By contrast, the extremely dense companion disc galaxies studied by \citeauthor{Mayer:2025aa} (\citeyear{Mayer:2025aa}, purple stars) at $z \sim 7.6$ reside within the halo of a much more massive galaxy, placing them in a very different environment than the isolated \textsc{GigaEris} system. The main sample of the \textsc{Phoebos} simulation, which includes many subhaloes, appears to be more similar to these compact companions in terms of gas content. This will also be explored further in future work.

\section{Summary and outlook}\label{sec:conclusions}

In this work, we introduced the \textsc{Phoebos} cosmological volume simulation, a new high-resolution model of early galaxy formation designed to probe the physical processes shaping galaxies during cosmic dawn. This paper set out with two primary objectives: first, to demonstrate the capability of \textsc{Phoebos} in accurately reproducing key galaxy statistics at high redshift and second, to provide a comprehensive analysis of galaxy sizes at redshifts $z \gtrsim 8$ across a wide range of galaxy stellar masses.

We find that \textsc{Phoebos} achieves both of these goals. The simulation is remarkably successful in capturing key observables of galaxy formation at high redshift.  Without any redshift-dependent calibration or tuning to high-redshift data, \textsc{Phoebos} accurately reproduces the stellar mass function, stellar-to-halo mass relation, and sSFR at $z \gtrsim 8$, in close agreement with current observational constraints \citep{Merlin:2019aa,Morishita:2024aa, Miller:2024aa, Paquereau:2025aa, Shuntov:2025aa, yang:2025aa}. Furthermore, we analysed the halo size-mass relation at $z = 8$ and $z = 12$, and compared the results with observational data from \citet{Morishita:2024aa} and \citet{yang:2025aa}. The  simulation does predicts larger sizes for galaxies than observed, likely due to the difference between the stellar half-mass radius used in the simulation and the effective radius used in the observations. This discrepancy is consistent with the known tendency for lower-resolution simulations to overestimate galaxy sizes \citep{Bate:1997aa}. Despite this, the \textsc{Phoebos} simulation accurately reproduces the slope of the size-mass relation up to a stellar mass of $\sim$$10^{9.5}$~M$_{\sun}$, matching well with \citet{yang:2025aa}. In contrast, the \textsc{Thesan} simulation \citep{Shen:2024aa} fails to reproduce the correct slope, highlighting the accuracy of \textsc{Phoebos} and its stellar feedback model in capturing the key physical processes of early galaxy growth.

Additionally, the SFR-stellar mass relation found within \textsc{Phoebos} fits the observations at $z \gtrsim 8$ by \citet{Morishita:2024aa}. The simulation primarily probes less massive galaxies, with more massive ones missing due to the limited box size. To fully capture the range of galaxy masses and SFRs, future simulations should utilize larger box sizes, enabling a more comprehensive comparison, particularly at the high-mass end.

The strong agreement between \textsc{Phoebos} and the observations suggests that the physical model, particularly its treatment of of a relatively weak stellar feedback, captures the key aspects of early galaxy evolution. Specifically, the results point to efficient SF being essential at high redshift for the formation of sufficiently massive galaxies during this epoch. This faster build-up of stellar mass may help interpret several recent JWST observations that indicate a more accelerated formation of galaxies than previously anticipated \citep[e.g.][]{Ferreira:2022aa, Harikane:2023aa, Huertas:2023aa, Kartaltepe:2023aa, Ono:2023aa, Adams:2024aa, Ito:2024aa, Ormerod:2024, yang:2025aa, Shuntov:2025aa}. This is particularly evident when comparing the results of the \textsc{Thesan} simulation \citep{Kannan:2022aa,Shen:2024aa} with those of \textsc{Phoebos} at $z \gtrsim 8$.  The feedback modeled in \textsc{Thesan} includes momentum-driven stellar winds and metal enrichment, coupled with a two-phase ISM treated via an effective equation of state \citep{Springel:2003aa}, stellar feedback via momentum-driven winds, metal enrichment, and SMBH growth with quasar and radio mode feedback. In \textsc{Phoebos}, the weaker feedback processes result in a better match to the observed evolution of galaxy sizes (Figure~\ref{fig:RMobs}) and the stellar-to-halo mass relation (Figure~\ref{fig:SHMR}). As well as, producing SF histories of the haloes (Figures~\ref{fig:SFR} and~\ref{fig:sSFR}) that are consistent with observations and \textsc{Thesan}.

The weak effect of the blastwave feedback sub-grid model in the extremely high gas density conditions of galaxies at $z > 7$ is manifested in the patterns exhibited by the phase diagrams, which nearly miss a substantial warm/hot gas phase (see Figure~\ref{fig:ULR}). This is reminiscent of the ``feedback-free'' scenario proposed by \citeauthor{Dekel:2023aa} \citep[e.g.][]{Dekel:2023aa, Dekel:2024aa, Dekel:2004ab}, which is indeed the consequence of cooling times and dynamical times becoming shorter than the typical time span of SN explosions (a few Myr). Note that the latter time-scale is independent on the specific form of SN energy or momentum coupling with the surrounding medium, hence it is captured correctly by a relatively simple model such as the blastwave feedback, in which there is no attempt to include additional physics behind the coupling, such as thermal conduction or energy dissipation in a multi-phase medium, or the collective effect of SN explosions. These additional aspects of feedback, which are tentatively included in more modern feedback recipes, would introduce additional time-scales, which apparently are not relevant to explain the high-redshift observations. Likewise, additional components of stellar feedback, such as photoionizing radiation from massive stars and radiation pressure, which are neglected in our sub-grid recipes, and also not fully taken into account in the proposed feedback-free model, would also add significant nuisances to the thermodynamics, which, it would seem, are also not important to explain the observables. Understanding this in detail, and not simply at a phenomenological level, will require radiative transfer calculations of individual star-forming regions in such high-redshift galaxies. In fact, on one hand gas collapse must proceed on a very fast track, which implies fast cooling times at least up to reaching a density at which gas becomes \citeauthor{Jeans1902}-unstable, apparently captured by our sub-grid models, while on the other the ionizing UV radiation must have long enough diffusion time to not prevent further collapse of nearby clouds, but also not produce too much radiation pressure to the detriment of gravitational collapse, in order for highly efficient SF to be sustained for long time-scales. We also note that the feedback-free regime, at the redshifts that are relevant here ($z \gtrsim 8)$, is expected to operate in galaxies hosted in haloes with virial mass of approximately $10^{10}$~M$_{\sun}$ and above \citep{Dekel:2023aa, Li:2024aa}. This is true for the most massive galaxies in our sample, but not for the majority of the galaxy population. The majority of galaxies should thus be in a regime in which feedback, although in a weak mode, is still acting to regulate SF. This is indeed consistent with the fact that the phase diagrams do not comprise only a cold gas phase, which would have been the case in a strictly ``feedback-free'' scenario.

In contrast, at later times such as $z \lesssim 2$, one expects the need for such weak feedback to lessen \citep[see, e.g. the simulation results of][]{Dubois:2014aa, kaviraj:2017aa, Nelson_et_al_2018, Naiman:2018aa, Marinacci:2018aa,  Nelson:2019aa, Dubois:2021aa, Schaye_et_al_2023, Oku:2024aa}. For example, simulations like TNG100 \citep[e.g.][]{Nelson_et_al_2018, Nelson:2019aa, Pillepich:2018aa, Marinacci:2018aa, Naiman:2018aa} and Horizon-AGN \citep[][]{Dubois:2014aa, kaviraj:2017aa} successfully reproduce the SHMR at $z \lesssim 1$, but tend to show limited evolution at higher redshifts. In particular, they often fail to capture the slope changes observed in recent JWST data \citep[see, e.g.][]{Paquereau:2025aa}. As shown in the top-right panel of Figure~\ref{fig:simprops}, it is likely that \textsc{Phoebos} will overshoot the cosmic SFR at these lower redshifts, suggesting a transition in the dominant physical processes governing galaxy growth. Understanding the nature of this transition, when it would happen, and its impact on the build up of stellar mass, should be a major objective for future theoretical and observational efforts (see Appendix~\ref{appen:phase} for an initial discussion based on the phase diagrams).

In addition to its weaker stellar feedback, the success of \textsc{Phoebos} may also come from its more realistic treatment of the ISM. Unlike many large-scale cosmological simulations that use an effective equation of state to model dense gas \citep{Springel:2003aa}, \textsc{Phoebos} makes more of an effort to capture the non-equilibrium cooling and the clumpy structure of the cold ISM. By resolving the clumpy, multiphase structure of the cold ISM, \textsc{Phoebos} provides a refined framework for galactic feedback processes. This improved resolution of the ISM’s complexity is crucial for capturing realistic SF and energy exchange in the early Universe. Notably, even simulations like \textsc{Thesan-Zoom} \citep{Kannan:2025aa}, which feature significantly higher resolution and stronger feedback, show results consistent with \textsc{Phoebos} and high-redshift observations \citep[e.g.][]{Morishita:2024aa, Harvey:2025aa}, highlighting the fundamental importance of detailed ISM physics. While this may indicate that the treatment of the ISM is more critical than the exact strength or implementation of stellar feedback, the current resolution and redshift limits of our simulations make it difficult to disentangle their relative contributions. Both components likely play complementary roles, and higher-resolution studies at both high and low redshifts will be necessary to clarify their individual impact on early galaxy formation.

Lastly, the results discussed on the HI fraction show that the \textsc{Phoebos} simulation will be able to offer valuable insights into HI gas content, already revealing a strong correlation between HI gas fraction and SF at high redshift. The simulation highlights the trend of more massive galaxies having lower HI fractions, which is consistent with observations from local surveys and simulated results. These findings will provide an important framework for future HI surveys and deepen our understanding of how gas content evolves with stellar mass and SF across cosmic time.

The success of \textsc{Phoebos} in the high-redshift regime offers a compelling framework for interpreting early galaxy formation. Its ability to reproduce key observables without redshift-specific tuning, combined with its physically motivated treatment of feedback and cooling, highlights the potential of this weaker feedback model and detailed treatment of the ISM to advance our understanding of galaxy formation across cosmic time.

\section*{Acknowledgements}
LM and FvD acknowledge support from the Swiss National Science Foundation under the Grant 200020\_207406. PRC and FvD acknowledge support from the Swiss National Science Foundation under the Sinergia Grant CRSII5\_213497 (GW-Learn). DSR acknowledges support from the Swiss State Secretariat for Education, Research and Innovation (SERI) through the SKACH grant.  We are grateful for computing grants provided by SKACH on Piz Daint at the Swiss National Supercomputing Centre (CSCS). We acknowledge Michael Tremmel for insightful discussions. We thank Takahiro Morishita for sharing their observational data with us.  

\section*{Data Availability Statement}
The data underlying this article will be shared on reasonable request to the corresponding author.

\scalefont{0.94}
\setlength{\bibhang}{1.6em}
\setlength\labelwidth{0.0em}
\bibliographystyle{mnras}
\bibliography{phoebos}
\normalsize

\newpage
\appendix
\section{Lower-resolution runs}
\begin{figure}
    \centering
    \includegraphics[ trim={0cm 0cm 0cm 0cm}, clip, width=0.48\textwidth, keepaspectratio]{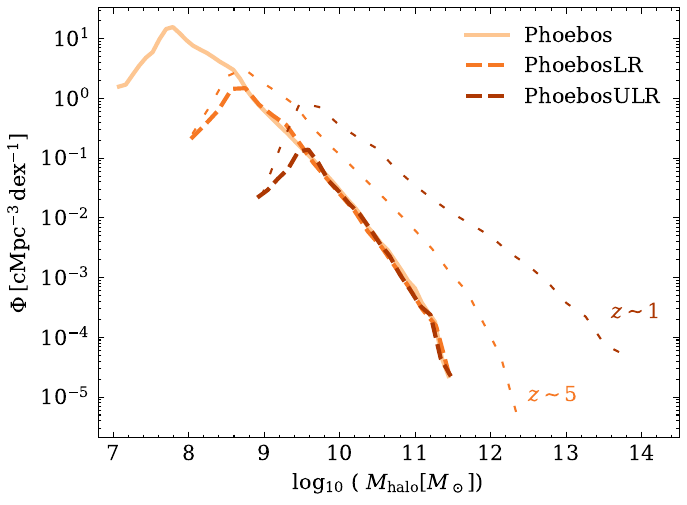}
    \caption{Halo mass functions across the three different runs described in this work at $z \sim 8$, corresponding to the final snapshot of the \textsc{Phoebos} run. The lower-resolution runs were evolved to lower redshifts: the halo mass function from the lowest-redshift snapshot is shown with loosely dashed lines.}
    \label{fig:halofunclow}
\end{figure}

As discussed in Section~\ref{sec:numerical_setup} and summarized in Table~\ref{tab:resolution}, the \textsc{Phoebos} simulation suite was run at three different resolution levels. Figure~\ref{fig:halofunclow} presents the resulting halo mass functions. The final snapshot at $z \sim 8$ from the highest-resolution run is shown alongside the lower-resolution runs, showing convergence. Additionally, we include the halo mass functions from the lowest redshift reached by each of the lower-resolution runs, shown with loosely dashed lines.

\subsection{Phase Diagram Evolution} \label{appen:phase}
\begin{figure}
    \centering
    \includegraphics[ trim={0cm 0cm 0cm 0cm}, clip, width=0.48\textwidth, keepaspectratio]{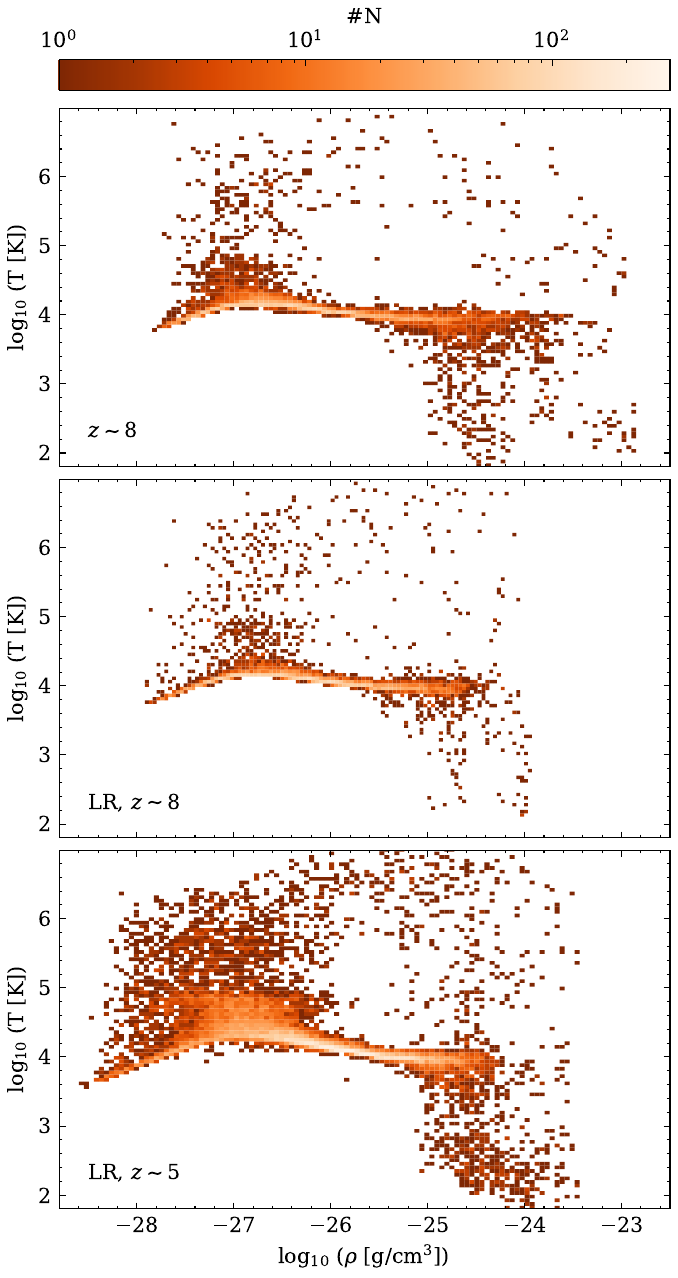}
    \caption{Gas phase diagrams of the 715th most massive galaxy at two redshifts in the \textsc{PhoebosLR} simulation. The bottom two panels show the temperature–density distribution of gas at $z \sim 8$ and $z \sim 5$, respectively. For comparison, the top panel shows the same quantity for the same halo at $z \sim 8$ from the higher-resolution \textsc{Phoebos} run.}
    \label{fig:ULR}
\end{figure}

\begin{figure}
    \centering
    \includegraphics[ trim={0cm 0cm 0cm 0cm}, clip, width=0.48\textwidth, keepaspectratio]{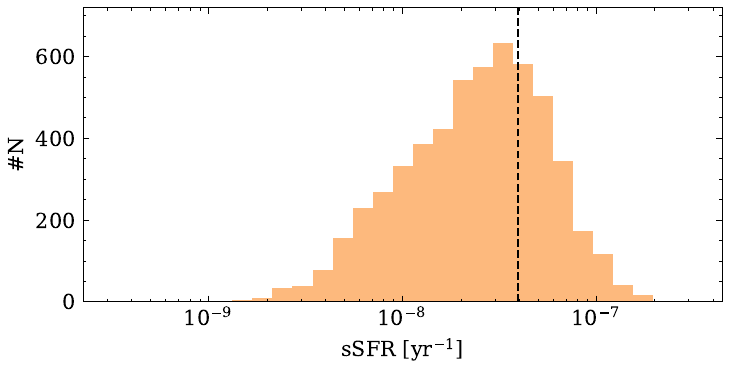}
   \caption{The sSFR distribution at $z= 5$ within the \textsc{PhoebosLR} run. The dashed black line indicates the halo chosen for the bottom panel in Figure~\ref{fig:ULR}.}
    \label{fig:SFRLRz5}
  \end{figure}

Figure~\ref{fig:ULR} presents the gas temperature–density phase diagrams for a representative galaxy, the 715th most massive at  $z \sim 5$, from the \textsc{PhoebosLR} simulation. The lower panel shows the galaxy at $z \sim 5$, while the upper two panels show the same halo at $z \sim8$ in both the high-resolution \textsc{Phoebos} run and the lower-resolution \textsc{PhoebosLR} run, demonstrating reasonable convergence across resolutions. This galaxy was selected for its high specific sSFR, as highlighted in Figure~\ref{fig:SFRLRz5}.

The gas is primarily concentrated around $10^4$ K at $z \sim 8$, with a negligible warm/hot gas component. This is consistent with a regime of rapid gas inflow and efficient radiative cooling, where cold accretion dominates and stellar feedback is largely ineffective. The lack of significant heating is indicative of a "feedback-free" or weak-feedback mode, as discussed in Section~\ref{sec:conclusions} and by \citet{Dekel:2023aa}, where short cooling and dynamical timescales prevent SN feedback from impacting the ISM thermodynamics. This leads to abundant star-forming gas and elevated SFRs, in line with observational trends at high redshift \citep[e.g.][]{Morishita:2024aa}.

By $z \sim 5$, the phase diagram shows the emergence of a substantial hot gas phase, especially at lower densities. While our sub-grid model does not include strong feedback mechanisms, this change may be partially attributed to the suppression of radiative cooling in low-density gas in the simulation, rather than to direct feedback heating. Nonetheless, this evolution reflects a transition toward a more regulated star-forming environment, with gas cycling becoming less efficient and the ISM developing a broader thermodynamic structure.

Interestingly, the emergence of warm ($\sim 10^5$ K), dense gas in the $z \sim 5$ panel echoes results from simulations with explicitly strong feedback implementations \citep[e.g.][]{Hopkins:2012aa, Fiacconi:2017aa}, suggesting that even relatively simple feedback prescriptions can qualitatively reproduce key features of ISM structure in evolving galaxies.

\begin{figure}
    \centering
    \includegraphics[ trim={0cm 0cm 0cm 0cm}, clip, width=0.48\textwidth, keepaspectratio]{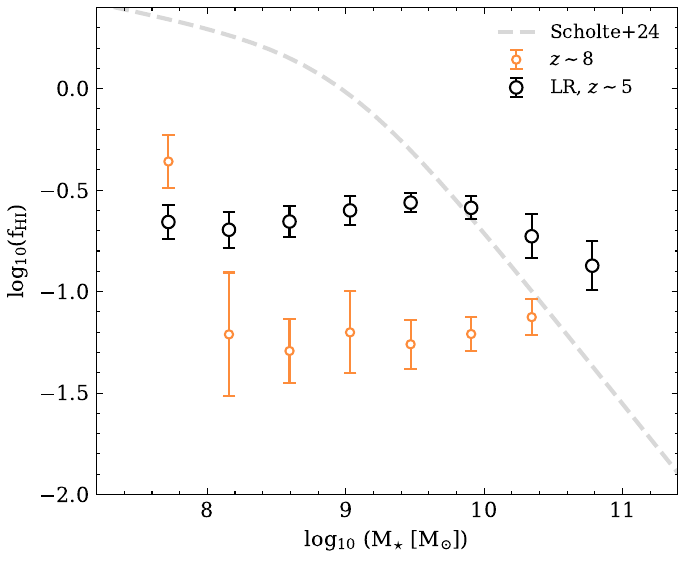}
   \caption{The stellar mass-HI gas fraction  relation at $z \sim 8$ for \textsc{Phoebos} and at $z \sim 5$ for \textsc{PhoebosLR}. The open circles and error bars represent the median and $1\sigma$ scatter of the HI gas fraction of the galaxies. }
    \label{fig:HILR}
  \end{figure}
  
\subsection{HI evolution}\label{appen:HI}

\begin{figure*}
    \centering
    \includegraphics[ trim={0cm 0cm 0cm 0cm}, clip, width=0.99\textwidth, keepaspectratio]{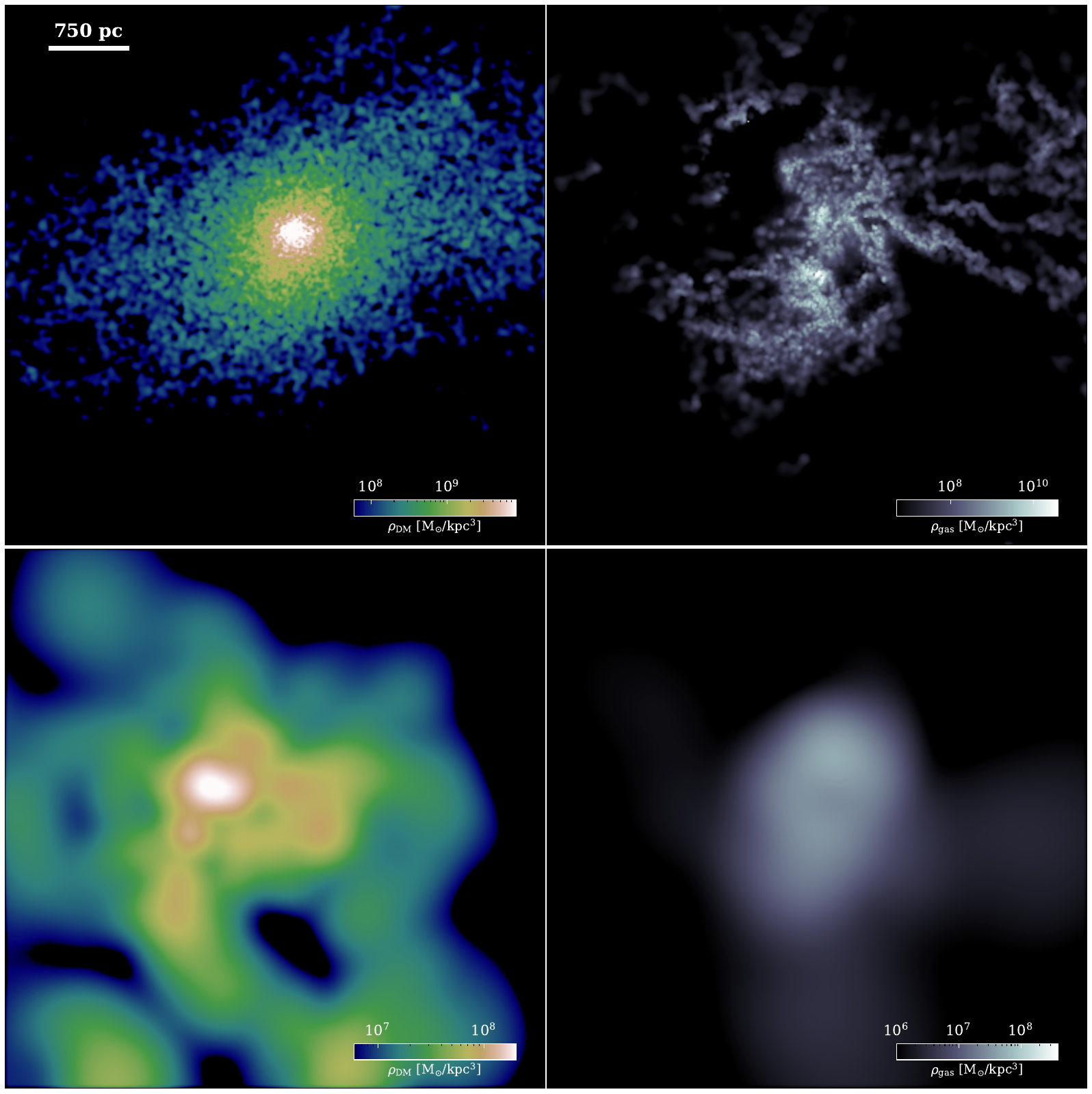}
    \caption{Comparison of the main galaxy in the \textsc{GigaEris} simulation (top panels) and a galaxy with a similar stellar and halo mass within \textsc{Phoebos} (bottom panels). The left-hand panels show the DM volume density distribution, whereas the right-hand panels display the corresponding gas volume density at $z \sim 8$. The galaxy within \textsc{Phoebos} has a halo (stellar) mass of $10^{9.67}$ ($10^{7.11}$)~M$_{\sun}$.}
    \label{fig:gigacomp}
\end{figure*}

To complement the main results, we show the $M_{\rm{HI}}$--$M_\star$ relation at $z \sim 5$ from the \textsc{PhoebosLR} run in Figure~\ref{fig:HILR}. This allows us to examine whether HI fractions increase towards lower redshifts. We find a modest increase in $f_{\rm{HI}}$ at fixed $M_\star$, consistent with a scenario where our feedback model becomes more effective at regulating SF with time. As a result, less HI is consumed in SF at lower redshift. 

It is important to note that this trend does not imply that the feedback model is universally efficient across all epochs, rather, the evolution of $f_{\rm{HI}}$ arises naturally within the simulation. However, as shown in the upper-left panel of Figure~\ref{fig:simprops}, \textsc{Phoebos} tends to overshoot the cosmic star formation rate density, indicating the need for stronger feedback at lower redshifts. Although achieving quantitative agreement at high redshift will ultimately require observational constraints from SKA data, demonstrating this trend already provides valuable insight.

We emphasize that our analysis focuses on global properties, such as the total HI mass fraction in galaxies. To fully understand how $f_{\rm{HI}}$ responds to feedback on smaller scales will necessitate higher-resolution studies \citep[e.g.][]{Gensior:2024aa}.

\section{Comparison to \textsc{GigaEris}}\label{appen:gig}

Figure~\ref{fig:gigacomp} presents a visual comparison between a galaxy from the \textsc{Phoebos} simulation and the main galaxy of the high-resolution zoom-in \textsc{GigaEris} simulation, both at $z \sim 8$. Although the two galaxies have comparable stellar and halo masses, \textsc{GigaEris} offers significantly higher resolution, capturing finer structural details and clearer features. Despite this, both simulations exhibit similar large-scale behavior, notably the gas inflows that feed the central galaxy.

\bsp 
\label{lastpage}
\end{document}